\documentclass{article}

\usepackage{microtype}
\usepackage{graphicx}
\usepackage{subcaption}
\usepackage{booktabs} 

\usepackage{hyperref}

\usepackage{algorithm}
\usepackage{algpseudocode} 
\usepackage{xcolor}        
\algrenewcommand\algorithmiccomment[1]{\hfill\textcolor{blue}{$\triangleright$~#1}}

\usepackage[preprint]{icml2026}

\usepackage{amsmath}
\usepackage{amssymb}
\usepackage{mathtools}
\usepackage{amsthm}

\usepackage[capitalize,noabbrev]{cleveref}

\theoremstyle{plain}

\theoremstyle{definition}

\theoremstyle{remark}

\usepackage[textsize=tiny]{todonotes}

\usepackage{stfloats}
\usepackage{changepage}
\usepackage{multirow}

\usepackage{placeins}

\usepackage{siunitx}
\usepackage[most]{tcolorbox}   
\usepackage{fontawesome5}            
\newtcolorbox{myprompt_single}[2][]
{
    breakable,
    colframe=black,      
    colback=gray!20,     
    boxrule=1pt,         
    arc=4pt,             
    left=10pt,           
    right=10pt,
    top=10pt,            
    bottom=10pt, 
    title=#2,
    #1
}

\newtcbox{\mybox}[1][red]{
    on line,
    breakable,
    arc = 0pt,
    outer arc = 0pt,
    colback = #1!10!white,
    colframe = #1!50!black,
    boxsep = 0pt,
    left = 0.5pt,
    right = 0.5pt,
    top = 0.1pt,
    bottom = 0.1pt,
    boxrule = 0pt,
    bottomrule = 1pt,
    toprule = 1pt
}

\newtcolorbox{myboxpro}[1][red]{%
  breakable,               
  arc=0pt, outer arc=0pt,
  colback=#1!10!white,
  colframe=#1!50!black,
  boxsep=0pt,
  left=1pt,
  right=1pt,
  top=2pt,
  bottom=2pt,
  boxrule=0pt,
  toprule=1pt,
  bottomrule=1pt,
  parbox=false,            
  before upper={\parindent=0pt}, 
}

\newtcolorbox{infopanel}[4][]{       
  enhanced,
  colback  = #2!white!1,
  colframe = #2,                     
  boxrule  = 1pt,
  arc      = 4pt,
  left     = 10pt,
  right    = 10pt,
  top      = 10pt,
  bottom   = 10pt,
  fonttitle = \bfseries\large,
  coltitle  = white,
  colbacktitle = #2!white!75,
  attach boxed title to top center={yshift=-2mm},
  title = {\faIcon{#3}\;#4},   
  #1                               
}

\icmltitlerunning{IGMiRAG: Intuition-Guided Retrieval-Augmented Generation with Adaptive Mining of In-Depth Memory }

\begin{document}
\twocolumn[
  \icmltitle{IGMiRAG: \underline{I}ntuition-\underline{G}uided Retrieval-Augmented\\ Generation with Adaptive \underline{Mi}ning of In-Depth Memory }



  \icmlsetsymbol{equal}{*}

  \begin{icmlauthorlist}
    \icmlauthor{Xingliang Hou}{se}
    \icmlauthor{Yuyan Liu}{se}
    \icmlauthor{Qi Sun}{se}
    \icmlauthor{Haoxiu Wang}{xh}
    \icmlauthor{Hao Hu}{icir}
    \icmlauthor{Shaoyi Du}{icir}
    \icmlauthor{Zhiqiang Tian}{se}
  \end{icmlauthorlist}
  
  \icmlaffiliation{se}{School of Software Engineering, Xi’an Jiaotong University}
  \icmlaffiliation{xh}{School of Engineering, Westlake University}
  \icmlaffiliation{icir}{State Key Laboratory of Human-Machine Hybrid Augmented Intelligence, National Engineering Research Center for Visual Information and Applications, and Institute of Artificial Intelligence and Robotics, Xi’an Jiaotong University}
  

  \icmlcorrespondingauthor{Zhiqiang Tian}{zhiqiangtian@xjtu.edu.cn}

  \icmlkeywords{LLMs, Retrieval-Augmented Generation, Hypergraph}

  \vskip 0.3in
]



\printAffiliationsAndNotice{}  



\begin{abstract}

Retrieval-augmented generation (RAG) equips large language models (LLMs) with reliable knowledge memory. To strengthen cross-text associations, recent research integrates graphs and hypergraphs into RAG to capture pairwise and multi-entity relations as structured links. However, their misaligned memory organization necessitates costly, disjointed retrieval.
To address these limitations, we propose IGMiRAG, a framework inspired by human intuition-guided reasoning. It constructs a hierarchical heterogeneous hypergraph to align multi-granular knowledge, incorporating deductive pathways to simulate realistic memory structures. During querying, IGMiRAG distills intuitive strategies via a question parser to control mining depth and memory window, and activates instantaneous memories as anchors using dual-focus retrieval. Mirroring human intuition, the framework guides retrieval resource allocation dynamically. Furthermore, we design a bidirectional diffusion algorithm that navigates deductive paths to mine in-depth memories, emulating human reasoning processes. Extensive evaluations indicate IGMiRAG outperforms the state-of-the-art baseline by $4.8\%$ EM and $5.0\%$ F1 overall, with token costs adapting to task complexity (average $6.3\,\mathrm{k}^{+}$, minimum $3.0\,\mathrm{k}^{+}$). This work presents a cost-effective RAG paradigm that improves both efficiency and effectiveness.

\end{abstract}

\begin{figure*}[ht]
  \begin{center}
    \centerline{\includegraphics[width=\textwidth]{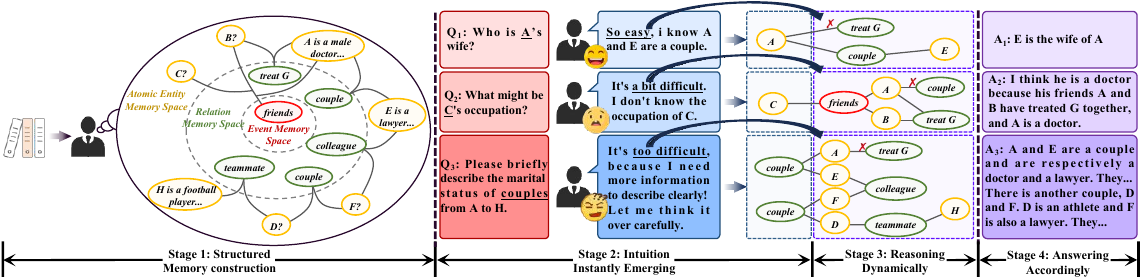}}
    \caption{
    $Stage\;1$ represents the formation of human memory, while $Stage\;2–Stage\;4$ represent intuition-guided reasoning mechanism in human cognition.
    Upon query encounter, we instantaneously assess and retrieve memory anchors, followed by deeper associative recall.
    }
    \label{human_intuition}
  \end{center}
\end{figure*}

\section{Introduction}
\label{Introduction}

Humans instantly activate long-term memory and think rapidly to generate goal-directed solutions in complex environments. Our cognitive maturity is rooted in the continuous consolidation of memory and reasoning. While Large Language Models (\textbf{LLMs}) leverage vast parametric knowledge for broad competence \cite{naseer2024integrating,dagdelen2024structured,ullah2024challenges}, their rigid, static memory often causes hallucinations. This limits their reliability for high-precision and complex decision-making tasks in dynamic environments \cite{huang2025survey,jones2025ai}.

To enhance LLM expertise and reliability, Retrieval-Augmented Generation (\textbf{RAG}) has become a dominant approach by non-parametrically integrating external knowledge, enabling low-cost updates \cite{lewis2020retrieval,asai2024self}. However, its reliance on flat vector similarity limits deep reasoning by failing to capture complex memory connections \cite{xie2023adaptive,zhong2024assessing}. To address this, recent research \cite{edge2024local,huang2025retrieval} attempts to build structured knowledge repositories. Graph-enhanced methods \cite{tian2024graph,wang2025archrag} map pairwise relationships as edges to form semantic paths, while hypergraph-enhanced methods \cite{feng2025hyper,hu2025cog} further leverage hyperedges to model high-order concepts like processes and events to establish associative paths among multiple entities.

Although these methods have built structured memory banks with semantic links, their homogeneous modeling often isolates basic and high-order knowledge as structurally disparate nodes and edges. This architectural misalignment leads to costly, fragmented retrieval, resulting in poor multi-hop reasoning and generalization \cite{jimenez2024hipporag}. Therefore, recent work \cite{xu2025noderag,gutierrez2025rag} employs heterogeneous graphs to align memories within a unified relevance space to mitigate this limitation, but the resulting connections remain limited to binary relations and lack systematic modeling.
Consequently, such disordered, low-order connectivity fundamentally limits the depth of memory association and reasoning.

How can RAG's memory retrieval be optimized without significant computational overhead? Inspired by human intuition-guided reasoning, we propose \textbf{IGMiRAG}, a novel framework featuring a more authentic memory architecture.
As illustrated in $Stage\,1$ of \cref{human_intuition}, the human brain organizes knowledge into a hierarchical associative network. This structure arises from inherent deductive pathways: high-order memories aggregate multiple low-order atomic facts, while single atoms contribute to various high-level concepts. Thus, knowledge is stored in a hierarchy that integrates multi-order associations.
Grounded in this, IGMiRAG emulates this structure using a Hierarchical Heterogeneous Hypergraph, where heterogeneous vertices represent multi-granular knowledge and hierarchical hyperedges model deductive pathways. This design establishes a foundation for efficient, human-like retrieval.


In response to specific tasks ($Stage\,2-Stage\,4$ in \cref{human_intuition}), humans rapidly activate experiential memories to form intuition, determining strategies and locating key memory anchors. Strategy-guided reasoning then proceeds along hierarchical deductive pathways. This association is inherently bidirectional: top-down (broadcast) activates subordinate low-order memories to enrich details, whereas bottom-up (screening) integrates shared high-order memories to distill abstraction. This enables efficient, in-depth reasoning within hierarchical repositories, facilitating precise decisions.
Inspired by this, IGMiRAG introduces a two-stage retrieval paradigm. First, it analyzes the query to generate an intuitive strategy, guiding the entire retrieval process. Subsequently, it executes preference-aware bidirectional diffusion based on activated anchors, thereby mining deep memories. Our main contributions are as follows:

\begin{itemize}
    \item \textbf{Better Memory Architecture}: We propose a Hierarchical Heterogeneous Hypergraph to model multi-granular knowledge, capturing deductive associations via hierarchical hyperedges. This architecture emulates human memory hierarchy, enabling interpretable and efficient retrieval.
    \item \textbf{Efficient Retrieval Paradigm}: We introduce an intuition-inspired ``Strategy-Diffusion'' two-stage retrieval paradigm. By first generating an intuitive strategy to guide the depth and scope of memory access, followed by anchor-based associative diffusion, this paradigm significantly improves both precision and efficiency with lower retrieval cost.
    \item \textbf{In-Depth Reasoning Mechanism}: We design a preference-aware bidirectional diffusion algorithm. By performing adaptive mining from anchors along the deductive pathways, it integrates top-down detail enrichment and bottom-up abstraction to support deep reasoning and significantly improve decision accuracy.
\end{itemize}



Evaluations on six benchmarks show IGMiRAG outperforms the state-of-the-art baseline by $4.8\%$ EM and $5.0\%$ F1 on average, with token costs adapting to task difficulty.
These results validate that by mimicking intuition-guided reasoning, our approach improves efficiency while simultaneously enhancing memory precision and reasoning depth, offering a viable solution to memory fragmentation and retrieval inefficiency in RAG systems.

\section{Related Works}
\textbf{Structure Optimization.}
Text-based RAG methods \cite{lewis2020retrieval,gao2023precise} extend LLM capabilities by simply concatenating raw text chunks, but often struggle with semantic sparsity and cross-contextual tasks \cite{gupta2024comprehensive}. Recent work has introduced graphs to systematically capture entity-level relationships, enhancing the richness of knowledge connections. However, these methods overlook higher-order multi-entity interactions, leading to information gaps \cite{srinivasan2018quantifying,santos2022knowledge,labatut2019extraction}. GraphRAG \cite{edge2024local} thus supplemented thematic summaries with dense community reports to enable macroscopic analysis.
Hypergraphs extend graphs by enabling a single hyperedge to connect multiple vertices at once \cite{gao2022hgnn+,feng2024beyond}. leveraging this capability, Hyper-RAG \cite{feng2025hyper} further unifies high-order multi-entity relations, thereby reducing fragmentation.
Nevertheless, both homogeneous graphs and hypergraphs only capture surface-level semantic links. While NodeRAG \cite{xu2025noderag} introduced node-type heterogeneity to encode cross-granular structural connections, IGMiRAG advances further by constructing a hierarchical heterogeneous hypergraph whose layered hyperedges explicitly encode deductive pathways, achieving improvements at both structural and semantic levels.

\textbf{Query Optimization.}
Beyond structural enrichment, a complementary line of work optimizes the query side to improve recall. Query rewriting narrows semantic gaps via context augmentation or rephrasing \cite{gao2023retrieval}, while query decomposition iteratively retrieves sub-answers through progressive queries to construct final answers \cite{chen2025you}. HyDE \cite{gao2023precise} generated hypothetical documents to enhance dense retrieval. Furthermore, keywords are also extracted to provide term-based matching with methods like BM25 \cite{robertson1994some}.
Although these methods have made improvements over static searchable sources, they still employ fixed strategies that lack task adaptivity.
IGMiRAG departs from this paradigm by parsing intuitive strategies from queries to jointly adapt the retrieval process, thereby enabling adaptive deep mining.

\textbf{Retrieval Optimization.}
Beyond these optimizations, some studies enhance retrieval effectiveness to obtain more relevant knowledge.
LightRAG \cite{guo2024lightrag} employs bidirectional expansion to augment adjacent knowledge. PathRAG \cite{chen2025pathrag} applies streaming pruning to suppress redundant paths. Cog-RAG \cite{hu2025cog} adopts a cognitive two-stage retrieval to ensure thematic consistency. The PPR \cite{haveliwala2002topic} algorithm is employed to mine multi-hop nodes \cite{xu2025noderag}. Despite these refinements, they are still constrained in multi-hop reasoning.
Guided by strategies, IGMiRAG drives a depth-adaptive process via deductive pathways to mine in-depth knowledge, significantly advancing multi-hop performance.


\section{IGMiRAG}

\begin{figure*}[ht]
  \begin{center}
  \includegraphics[width=\textwidth]{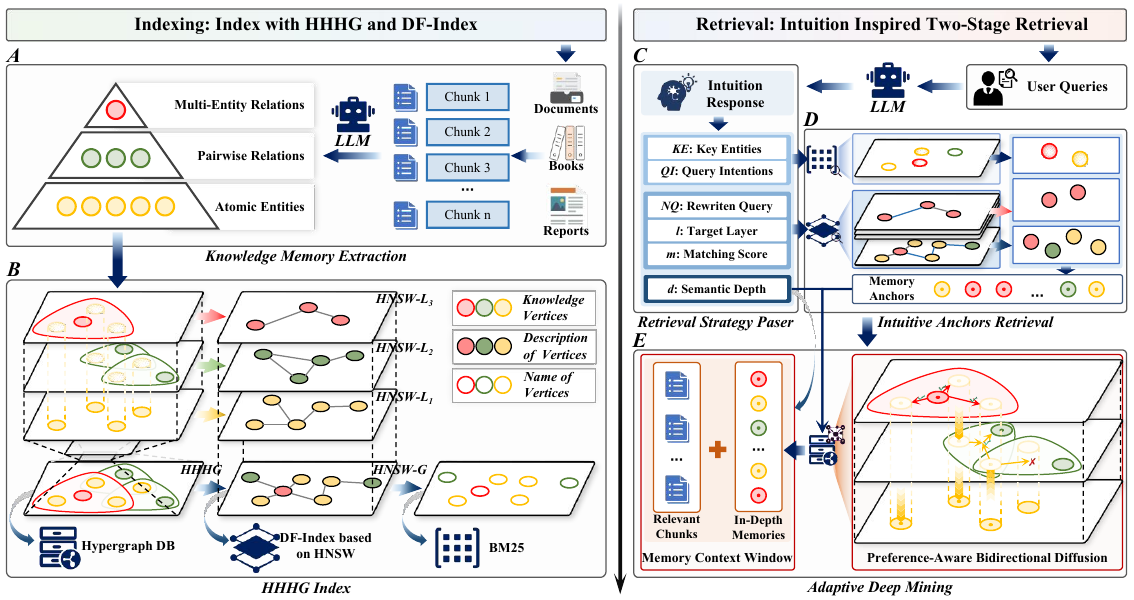}
  \caption{%
    \textbf{The framework of IGMiRAG.}
    \textbf{Indexing:}
    \textit{A}.~An LLM-based analyzer extracts multi-granular knowledge memories from each chunk.
    \textit{B}~Organizing all knowledge into a hierarchical heterogeneous hypergraph (HHHG) persisted in HyperGraph-DB. The semantic descriptions of all units are embedded and indexed as a global--local dual-focus HNSW index (DF-Index), and a separate BM25 corpus is built from name fields.
    \textbf{Retrieval:}
    \textit{C}.~An LLM-based Retrieval-Strategy Parser (RSP) distills strategies from user queries by simulating human intuition response.
    \textit{D}.~Multi-channel recall, combining BM25 string matching with dual-focus vector retrieval, identifies high-quality seed vertices as intuitive memory anchors.
    \textit{E}.~Preference-aware bidirectional diffusion traverses the HHHG to mine latent, in-depth memories. These units are then aggregated into a context window, the size of which is dynamically scaled according to query complexity, before being fed to the LLM.
    Two examples of the indexing and Retrieval process are provided in Appendix \ref{Pipeline_Example}.
  }
  \label{IGMiRAG_framework}
  \end{center}
\end{figure*}

\subsection{Overview}

As illustrated in \cref{IGMiRAG_framework}, IGMiRAG consists of four core components: HHHG Index, Retrieval-Strategy Parser (\textbf{RSP}), Intuitive Anchors Retrieval, and Adaptive Deep Mining.
Specifically, IGMiRAG employs an HHHG to encode knowledge memories and deductive pathways. Prior to retrieval, the RSP module distills intuitive strategies from queries, providing interpretable guidance for subsequent retrieval processes.
Subsequently, the framework first identifies high-quality knowledge as intuitive anchors via BM25 string matching and dual-focus retrieval.
During the deep mining phase, IGMiRAG performs preference-aware bidirectional diffusion along deductive pathways to uncover latent, relevant knowledge memories.
Finally, knowledge and chunks are adaptively selected based on semantic depth, which reflects the question complexity, yielding a compact yet highly relevant context that enables accurate and efficient knowledge-enhanced generation.


\subsection{Index construction}

\subsubsection{Hierarchical Heterogeneous Hypergraph}

By unifying four knowledge memory types and hierarchically modeling their deductive pathways, the HHHG architecture addresses the limitations of homogeneous graphs and hypergraphs in aligning diverse knowledge and cross-dimensional links simultaneously. Within IGMiRAG, an LLM-based analyzer processes raw chunks $\mathcal{D}$, encoding extracted entities $\mathcal{N}$, binary relations $\mathcal{L}$, and multi-entity relations $\mathcal{H}$ as dimensionally consistent heterogeneous vertices.
Deductive pathways are then encoded via hierarchical hyperedges, including $\mathcal{LR}$ ($\mathcal{N}\leftrightarrow\mathcal{L}$) and $\mathcal{HR}$ ($\mathcal{N}\leftrightarrow\mathcal{H}$). Formally as follows:
\begin{equation}
    \mathcal{G}_{\mathrm{HHHG}} = (\mathcal{V}, \mathcal{E}),
    \label{hypergraph}
\end{equation}
\begin{equation}
    \mathcal{V} = \mathcal{N} \cup \mathcal{L} \cup \mathcal{H},\quad
    \mathcal{E} = \mathcal{LR} \cup \mathcal{HR}.
    \label{V_E_set}
\end{equation}
Furthermore, $\mathcal{D}$ and $\mathcal{FR}$ ($\{\mathcal{N},\mathcal{L},\mathcal{H}\}\leftrightarrow\mathcal{D}$) serve exclusively to preserve source-text mapping.
The internal structures of $\mathcal{N}$, $\mathcal{L}$, and $\mathcal{H}$ are uniformly formalized as follows:
\begin{align}
    \mathcal{N} &= \bigl\{ \langle \mathit{Na}_i^{\mathcal{N}}, \mathit{De}_i^{\mathcal{N}}, \mathit{A}_i^{\mathcal{N}} \rangle \bigr\}_{i=1}^{|\mathcal{N}|}, \nonumber\\[-2pt]
    \mathcal{L} &= \bigl\{ \langle \mathit{Na}_j^{\mathcal{L}}, \mathit{De}_j^{\mathcal{L}}, \mathit{A}_j^{\mathcal{L}} \rangle \bigr\}_{j=1}^{|\mathcal{L}|}, \\[-2pt]
    \mathcal{H} &= \bigl\{ \langle \mathit{Na}_k^{\mathcal{H}}, \mathit{De}_k^{\mathcal{H}}, \mathit{A}_k^{\mathcal{H}} \rangle \bigr\}_{k=1}^{|\mathcal{H}|},\nonumber
    \label{NLH_set}
\end{align}
where $\mathit{Na}$, $\mathit{De}$, and $\mathit{A}$ denote the \emph{name}, \emph{semantic description}, and \emph{additional information} of knowledge memories respectively.
For $\mathcal{L}$ and $\mathcal{H}$, uniqueness is ensured by concatenating the associated entity names in a fixed order.

\subsubsection{Dual-Focus Index}
The Hierarchical Navigable Small World (\textbf{HNSW}) algorithm \cite{malkov2018efficient} is employed to construct a semantic-vector index for knowledge memories.
Notably, within the unified semantic space of the HHHG, single-pass global approximate nearest-neighbor retrieval is prone to cross-type semantic drift.
This occurs when vectorially proximate yet typologically irrelevant units introduce false positives, thereby degrading recall precision. 
To mitigate this drift, IGMiRAG introduces \textbf{DF-Index}, a dual-focus indexing library that operationalizes the intuitive focus within queries through local indexing over type-specific features.
The specific construction process is as follows:
\begin{equation}
    \mathcal{I}_{\mathrm{G}} = \mathrm{HNSW}(\{\mathit{De}^{\mathcal{N}}, \mathit{De}^{\mathcal{L}}, \mathit{De}^{\mathcal{H}}\}),
    \label{global_index}
\end{equation}
\begin{equation}
    \mathcal{I}_{\mathrm{L}} = \mathrm{HNSW}(\mathcal{X}),\quad
    \mathcal{X} \in \{\mathit{De}^{\mathcal{N}}, \mathit{De}^{\mathcal{L}}, \mathit{De}^{\mathcal{H}}\}.
    \label{local_index}
\end{equation}

\textbf{Global Indexing} inserts the \emph{semantic-description} vectors of all knowledge into a single HNSW graph $\mathcal{I}_{\mathrm{G}}$. This establishes a cross-type global navigation structure, enabling rapid localization of candidate regions across the entire knowledge space.
\textbf{Local Indexing} constructs separate HNSW subgraphs $\mathcal{I}_{\mathrm{L}}$ for vectors of each type in $\{\mathcal{N},\mathcal{L},\mathcal{H}\}$. 
Searching within these type-homogeneous neighborhoods could provide targeted supplements to the global candidates.


\subsection{Query Parsing}

User queries imply high-level strategic cues that integrate surface semantics with task-specific features (e.g., evaluative focus and response difficulty).
To exploit these cues, IGMiRAG employs an LLM-based RSP that simulates intuitive judgment. Beyond generating standard outputs, including the rewritten query ($\boldsymbol{\mathit{NQ}}$), key entities ($\boldsymbol{\mathit{KE}}$), and query intent ($\boldsymbol{\mathit{QI}}$), the RSP explicitly extracts two implicit signals. These signals provide fine-grained control over the subsequent retrieval and reasoning pathways.

\textbf{Target Layer $\mathit{l}$ and Matching Score $\mathit{m}$.}
The RSP predicts the knowledge layer $\mathit{l}\in\{\mathcal{N},\mathcal{L},\mathcal{H}\}$ most critical for answering and assigns a coverage score $\mathit{m}\in[1,5]\cap \mathbb{Z}$ reflecting the comprehensiveness of the query regarding that layer. The $(\mathit{l},\mathit{m})$ pair directs target-layer selection and modulates the global–local weighting during dual-focus retrieval.

\textbf{Semantic Depth $\mathit{d}$.}
The RSP analyzes $\mathit{NQ}$ to estimate complexity, abstraction, and inference depth, producing a semantic depth $d\in[1,5]\cap\mathbb{Z}$. The depth determines the number of diffusion iterations and the context-window size.


By decomposing the query into a multidimensional retrieval strategy, the RSP shifts retrieval from a passive, static process to an active, strategy-guided one, providing an interpretable and quantifiable basis for adaptive deep mining.


\subsection{Intuitive Anchors Retrieval}
\subsubsection{Multi-channel recall}

\textbf{Keywords Matching.}
Using the $\mathit{KE}$ and $\mathit{QI}$ produced by the RSP, IGMiRAG forms a composite query via term concatenation. BM25 matching is subsequently applied to the name field of all knowledge units, with the resulting candidates $\mathcal{C}_{\text{BM25}}$ ranked by descending BM25 score.

\textbf{Dual-Focus Vector Retrieval (DF-Retrieval).}
Beyond keyword matching, IGMiRAG employs a dual-focus semantic retrieval mechanism. Guided by the target layer $\mathit{l}$ and matching score $\mathit{m}$ provided by the RSP, the system searches the corresponding HNSW subgraph, thereby reinforcing global recall while suppressing cross-type semantic drift.

Given a base quota $k_b$, and the allowed bounds $k_{\text{min}}$, $k_{\text{max}}$ for global retrieval, and the dynamic quota for DF-Retrieval are defined as follows:
\begin{equation}
    k_G = \min\bigl(\left\lceil\left(1-\frac{m}{6}\right)\cdot k_b + k_{\text{min}}\right\rceil,\; k_{max}\bigl),
    \label{g_topk}
\end{equation}
\begin{equation}
    k_L = \left\lfloor\frac{m}{6}\cdot k_b\right\rfloor.
    \label{l_topk}
\end{equation}
The resulting candidates $\mathcal{C}_{\text{DF}}$ are sorted by descending similarity score.
Local rankings are appended subsequent to the global rankings, ensuring global candidates retain priority.
The union $\mathcal{C}_{\text{BM25}}\cup \mathcal{C}_{\text{DF}}$ finally constitutes intuitive anchors.

\subsubsection{RRF Fusion and Chunk Relevance}
Both rankings are converted to RRF scores \cite{cormack2009reciprocal} as relevance scores $s(v)$ with smoothing hyper-parameter $k_{0}=60$ :
\begin{equation}
s(v)=\sum_{c\in \{\mathcal{C}_{\text{BM25}}, \mathcal{C}_{\text{DF}}\}}\frac{1}{k_{0}+\text{rank}_{c}(v)},\; v\in \mathcal{C}_{\text{BM25}}\cup\mathcal{C}_{\text{DF}}.
\end{equation}
Scores $s(v)$ are propagated to chunks via unit–chunk associations.
After normalizing each $s(v)$ by its degree (number of associated chunks) to reduce bias from high-frequency vertices, scores are accumulated per chunk to produce the initial chunk-relevance score $s(c)$ as below:
\begin{equation}
s(c)=\sum_{(v,c)\in \mathcal{FR}}\frac{s(v)}{|\{c'\mid(v,c')\in \mathcal{FR}\}|},\;c\in \mathcal{D}.
\label{chunk_relevance}
\end{equation}

\subsection{Adaptive Deep Mining}
\subsubsection{Preference-Aware Bidirectional Diffusion}


Drawing on the human association introduced in \cref{Introduction}, three governing principles are posited:
(i)~vertex relevance is positively correlated with adjacency quality; (ii)~diffusion should be directionally amplified along deductive paths;
and (iii)~the required diffusion depth scales positively with question complexity.

Under these constraints, a \textbf{Preference-Aware Bidirectional Diffusion (PABD)} algorithm is formulated. It initiates from anchors and conducts a hierarchical, bidirectional diffusion process across the $\text{HHHG}$. 
The corresponding workflow and pseudocode are provided in Appendix \ref{PABD_Algorithm}.

\textbf{Propagation Mode.} Each iteration consists of two sequential phases: (i)~top-down broadcasting, in which high-order vertices propagate signals along $\langle\mathcal{H}\!\to\!\mathcal{N},\mathcal{L}\!\to\!\mathcal{N}\rangle$ to strengthen relevant lower-level memories; and (ii)~bottom-up screening, wherein low-order vertices feed signals back along $\langle\mathcal{N}\!\to\!\mathcal{H},\mathcal{N}\!\to\!\mathcal{L}\rangle$ only when co-occurrence support exceeds the adaptive threshold $\tau$, thereby reinforcing the corresponding higher-level memories.

\textbf{Preference-aware mechanism.} The preference coefficient~$\rho$ integrates the count of adjacent propagation sources with their normalized weights. This mechanism amplifies scores along high-preference paths while attenuating those in low-preference directions. Consequently, vertices reachable via deep multi-hop propagation overcome distance-based decay, achieving competitive relevance scores.


\textbf{Dynamic threshold.} The default thresholds are set to $\tau_0^{(\mathcal{L})}=0.5$ and $\tau_0^{(\mathcal{H})}=0.4$. A bias $b$ is automatically adjusted based on activation feedback during each iteration. If no new vertices are activated, $b$ increases to trigger backtracking; otherwise, it decreases. Ultimately, The effective threshold is $\tau^{(t)}=\tau_0^{(t)}-b$, where $t\in\{\mathcal{L},\mathcal{H}\}$. This mechanism balances exploration and exploitation, preventing diffusion stagnation while suppressing noise amplification to ensure controllable propagation.


The PABD algorithm terminates upon reaching the query’s semantic depth $d$ or upon diffusion stagnation. 
The activated vertices are sorted in descending order of extended relevance $s'(v)$ for context-window truncation.

\subsubsection{Adaptive Context-Window}

\textbf{Depth Amplification Mechanism.}
Defined by default as $k_{u}=5$ (knowledge-unit multiplier) and $k_{c}=2$ (chunk multiplier), these amplification coefficients jointly determine the adaptive recall quotas:
\begin{equation}
\mathrm{Top}\text{-}K_u=k_{u}\cdot d,\quad \mathrm{Top}\text{-}K_c=k_{c}\cdot d.
\end{equation}
\textbf{Final knowledge selection.}
After diffusion ends, the $\mathrm{Top}\text{-}K_u$ expanded vertices (excluding anchors) are extracted from the PABD output and merged with the initial vertices to form the final knowledge set $set_u$.

\textbf{Final chunks selection.}
The extended relevance score of each chunk is computed by applying the same normalized accumulation strategy as \cref{chunk_relevance} to the vertex-extension scores $s'(v)$.
The final chunk relevance is derived via a weighted fusion of these extended scores with the initial relevance values. Chunks are then ranked by this composite score in descending order. The $\mathrm{Top}\text{-}K_c$ selections, combined with the filtered $set_u$, constitute the retrieved content for the final context window.

The system utilizes the depth amplification coefficient and an ``initial–expansion'' fusion strategy to adaptively scale the memory window. This guarantees comprehensive knowledge coverage for complex problems while preserving efficiency and cost-effectiveness for simple queries.

\begin{table*}[hb]
    \begin{center}
        \caption{
        \textbf{QA performance} including EM, F1 scores (\%) on six RAG benchmarks. This table, along with the following ones, highlight the \textbf{best} and \underline{second-best} results.
        }
        \label{tab_main}

\begin{tabular}{lcc|cccccc|cccc|cc}
\toprule
    \multirow{3}{*}{\textbf{Methods}}
    & \multicolumn{2}{c}{\textbf{Simple QA}} 
    & \multicolumn{6}{c}{\textbf{Multi-Hop QA}} 
    & \multicolumn{4}{c}{\textbf{Explanatory QA}} 
    & \multicolumn{2}{c}{\textbf{Overall}} \\
    
\cmidrule(lr){2-3}\cmidrule(lr){4-9}\cmidrule(lr){10-13}\cmidrule(lr){14-15}

    & \multicolumn{2}{c}{PopQA} 
    & \multicolumn{2}{c}{MuSiQue} 
    & \multicolumn{2}{c}{2Wiki} 
    & \multicolumn{2}{c}{HotpotQA} 
    & \multicolumn{2}{c}{Mix} 
    & \multicolumn{2}{c}{Pathology} 
    & \multicolumn{2}{c}{Avg.}
    \\

 
\cmidrule(lr){2-3}\cmidrule(lr){4-5}\cmidrule(lr){6-7}\cmidrule(lr){8-9}\cmidrule(lr){10-11}\cmidrule(lr){12-13}\cmidrule(lr){14-15}
    & EM & F1
    & EM & F1
    & EM & F1
    & EM & F1
    & EM & F1
    & EM & F1
    & EM & F1
    \\
    
\midrule
    GPT-4o-mini & 20.7 & 24.5 & 12.4 & 21.7 & 32.2 & 37.4 & 30.8 & 41.4 & 57.5 & 57.5 & 72.1 & 69.4 & 37.6 & 42.0\\

\multicolumn{15}{l}{\textbf{\textit{Naive RAG}}}\\

    RAG\,($Top{-}1$) & 41.3 & 51.9 & 21.4 & 31.4 & 28.8 & 35.1 & 41.4 & 55.6 & 69.6 & 68.4 & 76.5 & 74.8 & 46.5 & 52.9\\
    RAG\,($Top{-}3$) & 46.6 & 58.6 & 27.4 & 38.4 & 37.9 & 45.1 & 48.5 & 62.7 & 73.9 & 72.2 & 76.8 & 75.5 & 51.8 & 58.8\\
    RAG\,($Top{-}5$) & 48.9 & 60.6 & \underline{28.6} & \underline{40.3} & 41.1 & 48.9 & \underline{50.2} & \underline{65.1} & 74.5 & 73.0 & 77.3 & 75.9 & 53.4 & 60.6\\

\multicolumn{15}{l}{\textbf{\textit{Graph-enhanced RAG}}}\\

    LightRAG & 49.1 & 60.3 & 22.6 & 32.4 & 35.2 & 41.4 & 40.3 & 54.3 & 74.5 & 72.8 & 75.5 & 73.6 & 49.5 & 55.8\\
    PathRAG & 33.1 & 44.3 & 19.3 & 31.1 & 40.2 & 47.8 & 41.1 & 55.8 & 56.8 & 56.1 & 61.9 & 61.8 & 42.1 & 49.5\\
    NodeRAG & \textbf{50.2} & \textbf{62.7} & 27.8 & 40.1 & 40.9 & 49.7 & 50.1 & 64.0 & \underline{75.3} & \underline{73.5} & 76.6 & 75.6 & \underline{53.5} & \underline{60.9}\\

\multicolumn{15}{l}{\textbf{\textit{Hypergraph-enhanced RAG}}}\\

Hyper-RAG & 49.1 & 59.9 & 22.6 & 32.6 & \underline{45.3} & \underline{54.0} & 41.0 & 54.9 & 71.4 & 69.7 & 75.9 & 75.2 & 50.9 & 57.7\\
Cog-RAG & 30.1 & 44.3 & 15.5 & 26.6 & 27.5 & 40.7 & 28.1 & 44.1 & 73.7 & 72.2 & \underline{78.9} & \textbf{78.0} & 42.3 & 51.0\\
\midrule
IGMiRAG (Ours) & \underline{49.8} & \underline{62.4} & \textbf{33.0} & \textbf{45.0} & \textbf{57.5} & \textbf{67.4} & \textbf{54.0} & \textbf{69.1} & \textbf{76.5} & \textbf{74.4} & \textbf{79.2} & \underline{77.4} & \textbf{58.3} & \textbf{65.9}\\
\bottomrule
\end{tabular}
    \end{center}
\end{table*}


\section{Experiment}
\subsection{Experimental Setup}
\textbf{Baselines.} 
We compared our approach with the state-of-the-art (SOTA) and widely adopted RAG methods. These include: Naive RAG; graph-enhanced methods ( LightRAG, PathRAG, and NodeRAG) and hypergraph-enhanced methods (Hyper-RAG and Cog-RAG). Detailed descriptions of these baselines are provided in Appendix\,\ref{Baselines_Details}.

\textbf{Benchmarks.} 
To comprehensively assess the cross-task generalisation of RAG methods, we utilize six public benchmarks spanning three representative tasks: (i)~\textbf{Detail Capture (Simple QA)}—PopQA \cite{mallen2023not}; (ii)~\textbf{Multi-Hop Reasoning (Multi-Hop QA)}—MuSiQue \cite{trivedi2022musique}and 2WikiMultihop \cite{ho2020constructing} (2Wiki), and HotpotQA \cite{yang2018hotpotqa}; and (iii)~\textbf{Knowledge Explanation (Explanatory QA)}—Mix \cite{qian2024memorag} and Pathology \cite{xiong2024benchmarking}.
Simple QA targets single-fact detail questions, and Multi-Hop QA demands cross-context aggregation and logical deduction, emphasizing error correction and chained reasoning following initial retrieval failures. Explanatory QA utilizes long passages with random-hop questions to evaluate comprehensive semantic representation and latent association mining.
Benchmark statistics are detailed in Appendix\,\ref{Benchmarks_Details}.

\textbf{Metrics.} We adopt three metrics across tasks: Exact Match (\textbf{EM}) and \textbf{F1} scores to measure literal overlap and n-gram recall rates between model outputs and reference answers, and average tokens per query (\textbf{Avg. Tokens}) to measure method efficiency.
For Explanatory QA, we employ an LLM-based evaluator to assign EM and F1 scores, assessing factual correctness and semantic relevance. Metric details are provided in Appendix\,\ref{Metrics_Details}.

\textbf{Implementation Details.} We employ text-embedding-3-small for text encoding and GPT-4o-mini \cite{achiam2023gpt} as the LLM. For IGMiRAG, the slice length is set to $1024$ tokens for Explanatory QA and $780$ for all others.
All baselines are configured using their officially recommended indexing and retrieval hyperparameters (detailed in Appendix\,\ref{Implementation_Details}) to ensure fair and reproducible comparisons.

\begin{figure*}[ht]
  \centering
  \begin{subfigure}[b]{0.65\textwidth}
    \begin{center}
        \includegraphics[width=\linewidth]{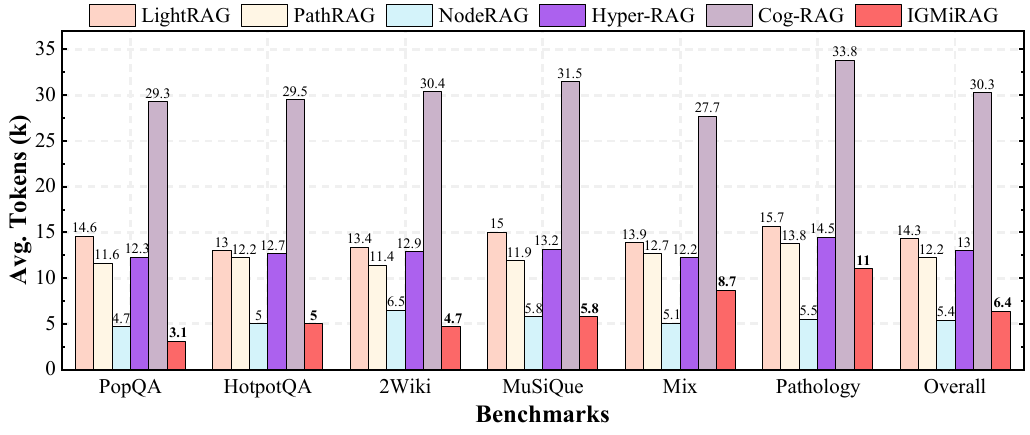}
        \caption{The comparison of average token costs}
        \label{avg_tokens}
    \end{center}
  \end{subfigure}\hfill
  \begin{subfigure}[b]{0.35\textwidth}
    \begin{center}
        \includegraphics[width=\linewidth]{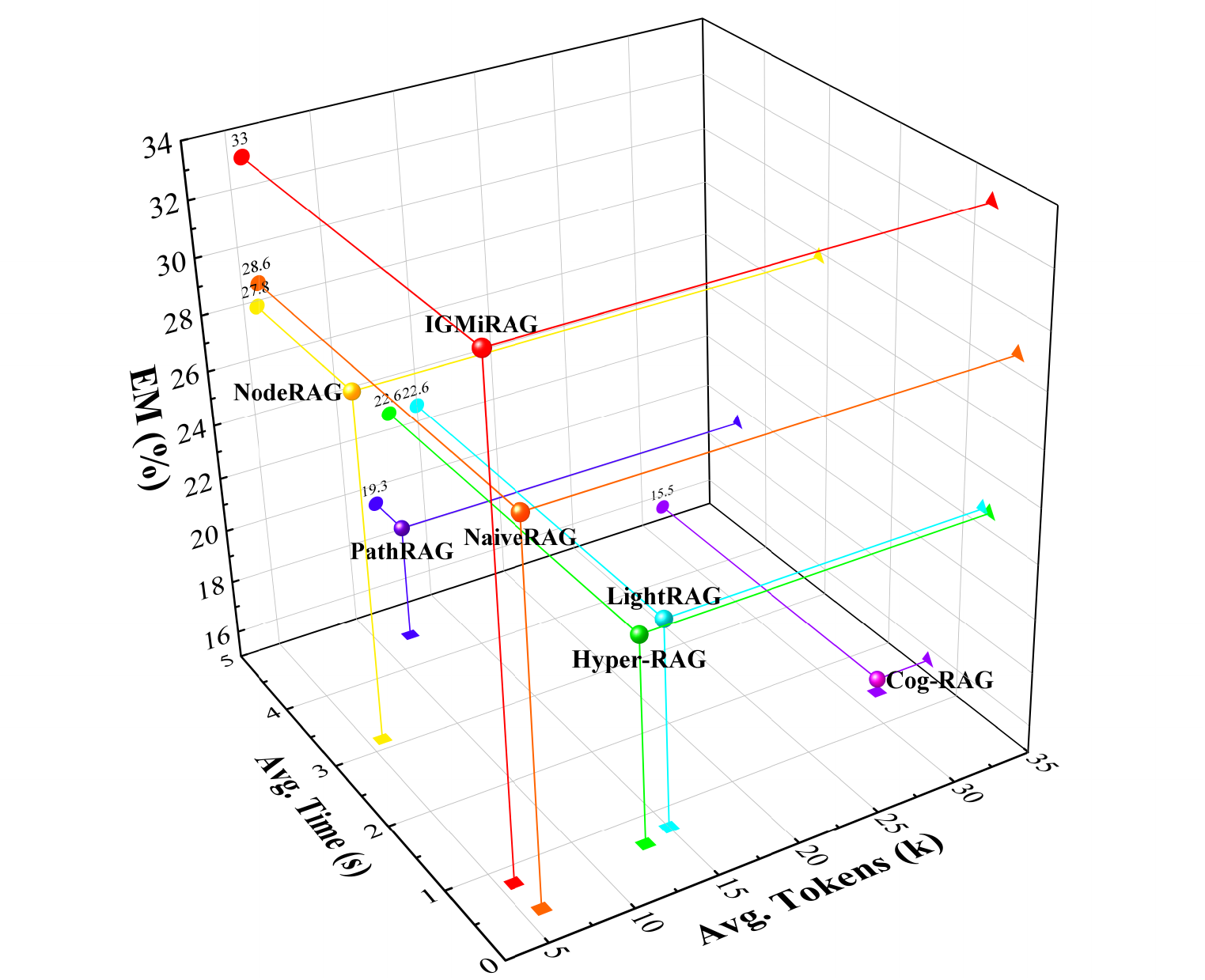}
        \caption{The comprehensive comparison}
        \label{overall_comparison}
    \end{center}
  \end{subfigure}
  \caption{
    \textbf{Efficiency Comparison.} Subfigure $(a)$ shows the token costs\,(k) comparison across structure-enhanced methods, while subfigure $(b)$ provides a comprehensive comparison of all RAG methods on MuSiQue regarding \textbf{EM}, \textbf{Avg.\ Tokens}, and \textbf{Avg.\ Time}.%
  }
  \label{efficiency_evaluation}
\end{figure*}


\subsection{Main Results}
\label{main_results}



We report QA performance and average token costs per query across all benchmarks, calculated against the gold-standard question–answer pairs.

\textbf{QA Performance.} \cref{tab_main} presents the performance of each method across six benchmarks. IGMiRAG achieves the highest average scores ($58.3\%$ EM, $65.9\%$ F1), maintaining a consistent lead across different tasks.
Naive RAG exhibits monotonic improvement with additional slices, albeit with diminishing marginal returns. While Naive RAG\,($Top{-}5$) ranks third overall, it performs second-best on MuSiQue and HotpotQA, outperforming several structure-enhanced methods despite its lower retrieval cost.
Among graph-enhanced methods, PathRAG ranks last overall ($42.1\%$ EM, $49.5\%$ F1), with minimal gains ($4.5\%$ EM and $7.5\%$ F1) over the LLM baseline. NodeRAG, however, emerges as the SOTA baseline, securing the top spot on Simple QA and second place on Mix.
Hypergraph-enhanced Cog-RAG observably exceeds Hyper-RAG on Explanatory QA but underperforms on Simple QA and Multi-Hop QA, with drops up to $19.0\%$ EM and $15.6\%$ F1 on PopQA.
Crucially, on the two most challenging benchmarks—MuSiQue and 2WikiMultiHop, IGMiRAG surpasses the second-best method by $4.4\%/4.7\%$ and $12.2\%/13.4\%$ in EM/F1, respectively, confirming its effectiveness in multi-hop reasoning. Besides securing second place on Simple QA, IGMiRAG achieves the highest performance on all five other benchmarks, demonstrating strong generalization and robustness.


\textbf{QA Efficiency.}
Query efficiency is critical for the practical viability of RAG systems. 
As shown in \cref{efficiency_evaluation}\,(\subref{avg_tokens}), the token consumption varies significantly across structure-enhanced methods. NodeRAG consumes the fewest tokens on average $5.4\,\mathrm{k}^{+}$, while IGMiRAG requires only $\approx$\,$0.9\,\mathrm{k}$ more ($6.3\,\mathrm{k}^{+}$). In contrast, all other methods exceed $11\,\mathrm{k}$ tokens (maximum $33.8\,\mathrm{k}^{+}$) per query.
A task-level breakdown reveals the following efficiency/performance trade-offs.
(i) Simple QA: IGMiRAG ranks second in performance while consuming only $3.0\,\mathrm{k}^{+}$ avg.\ tokens—$1.7\,\mathrm{k}^{+}$ fewer than NodeRAG ($4.7\,\mathrm{k}^{+}$).
(ii) Multi-Hop QA: Averaged across three benchmarks, IGMiRAG achieves the best performance with a minimum of $5.1\,\mathrm{k}^{+}$ tokens, $10.37\%$ ($0.5\,\mathrm{k}^{+}$) less than NodeRAG ($5.7\,\mathrm{k}^{+}$) and $60.04\%$ ($7.7\,\mathrm{k}^{+}$) less than Hyper-RAG ($12.9\,\mathrm{k}^{+}$).
(iii) For Explanatory QA, where cross-domain knowledge fusion is required, IGMiRAG increases its token usage to $8.6\,\mathrm{k}^{+}$ on Mix and $11.0\,\mathrm{k}^{+}$ on Pathology. Despite this increase, it achieves the highest performance at the second-lowest token cost.
Notably, the token consumption of all baselines, including NodeRAG, remains approximately constant across varying task complexities, exhibiting only minor and irregular fluctuations. In contrast, IGMiRAG's consumption scales proportionally with task complexity, enabling cost savings while maintaining strong performance.
Furthermore, \cref{efficiency_evaluation}\,(\subref{overall_comparison}) presents a comprehensive comparison of all RAG methods on MuSiQue. Obviously, IGMiRAG achieves the highest performance while maintaining both low time and token consumption.

\section{Discussions}
The following sections will analyze the effectiveness of each proposed module and its internal mechanisms, clarifying their individual contributions and synergistic interactions.





\begin{table}[ht]
  \caption{
    \textbf{Ablations.}
    We compare the QA performance and token cost of alternative retrieval and diffusion strategies on MuSiQue against the final IGMiRAG configuration.
    }
  \label{ablation_table}
  \begin{center}
    \begin{small}

        \begin{tabular}{lccc}
            \toprule
                \textbf{Models}
                & EM (\%)
                & F1 (\%)
                & Avg.\ (k)
                \\
            
            \midrule
            IGMiRAG & \textbf{33.0} & \textbf{45.0} & 5.84\\
            \midrule
            \multicolumn{4}{l}{\textbf{\textit{Retrieval Ablation}}}\\
            w/o BM25 & 30.6 & 41.58 & 5.60\\
            w/o DF-Retrieval\\
            \;\;\;\;w/o $\mathcal{I}_{\mathrm{L}}$ & 28.1 & 41.0 & \underline{5.56}\\
            \;\;\;\;w/ $\mathcal{I}_{\mathrm{G}}$ ($Top{-}10$) & 29.5 & 42.6 & 5.65\\
            \;\;\;\;w/ $\mathcal{I}_{\mathrm{G}}$ ($Top{-}20$) & 31.6 & 43.2 & 6.56\\
            \;\;\;\;w/ $\mathcal{I}_{\mathrm{G}}$ (dynamic $Topk$) & 31.6 & 43.6 & 6.19\\
            \midrule
            \multicolumn{4}{l}{\textbf{\textit{Diffusion Ablation}}}\\
            w/o PABD & 31.3 & 42.5 & \textbf{4.44}\\
            \;\;\;\;w/o Dynamic Threshold & 31.7 & 44.1 & 5.84\\
            \;\;\;\;w/o Preference Aware & \underline{32.4} & \underline{44.4} & 5.75\\
            \bottomrule
        \end{tabular}

    \end{small}
  \end{center}
  \vskip -0.1in
\end{table}


\subsection{Ablation Study}
To validate the effectiveness of the strategy-driven retrieval mechanism, we conducted ablation studies on MuSiQue while keeping the question parsing strategy fixed. 
As shown in \cref{ablation_table}, each module and mechanism yields a significant performance improvement for IGMiRAG.

\textbf{Retrieval Ablation.}
Both BM25 and DF-Retrieval improve initial candidates quality, with the latter yielding larger gains. Ablating local semantic retrieval alone reduces EM by $4.1\%$ and F1 by $3.7\%$. Three global-only ablations further confirm this: while $\mathcal{I}_{\mathrm{G}}$ $Top{-}20$ outperforms $\mathcal{I}_{\mathrm{G}}$ $Top{-}10$, it matches the performance of Dynamic $\mathcal{I}_{\mathrm{G}}$ (which reallocates the local retrieval budget to the global while maintaining the same $Topk$ as DF-Retrieval) yet consuming $0.3\,\mathrm{k}^{+}$ extra tokens. 
This indicates that expanding global candidates introduces redundancy and inefficiency. In contrast, the local perspective injects targeted units atop the global candidates, effectively offsetting the diminishing returns of scale.

\textbf{Diffusion Ablation.}
Removing the PABD module results in answering based solely on initial retrieval, consuming only $4.4\,\mathrm{k}^{+}$ tokens. In contrast, dropping the dynamic threshold mechanism degrades diffusion into indiscriminate expansion, amplifying high-order noise with longer texts. Consequently, this ablation suffers a worse performance drop than removing the preference-aware mechanism alone, despite the higher costs. Crucially, across all ablations, degrading the diffusion process proves less detrimental than impairing retrieval quality. This confirms that low-quality seeds steer diffusion along false paths, ultimately amplifying errors.


\begin{figure}[ht]
\begin{center}
  \begin{subfigure}[b]{0.48\columnwidth}
    \begin{center}
        \includegraphics[width=\linewidth]{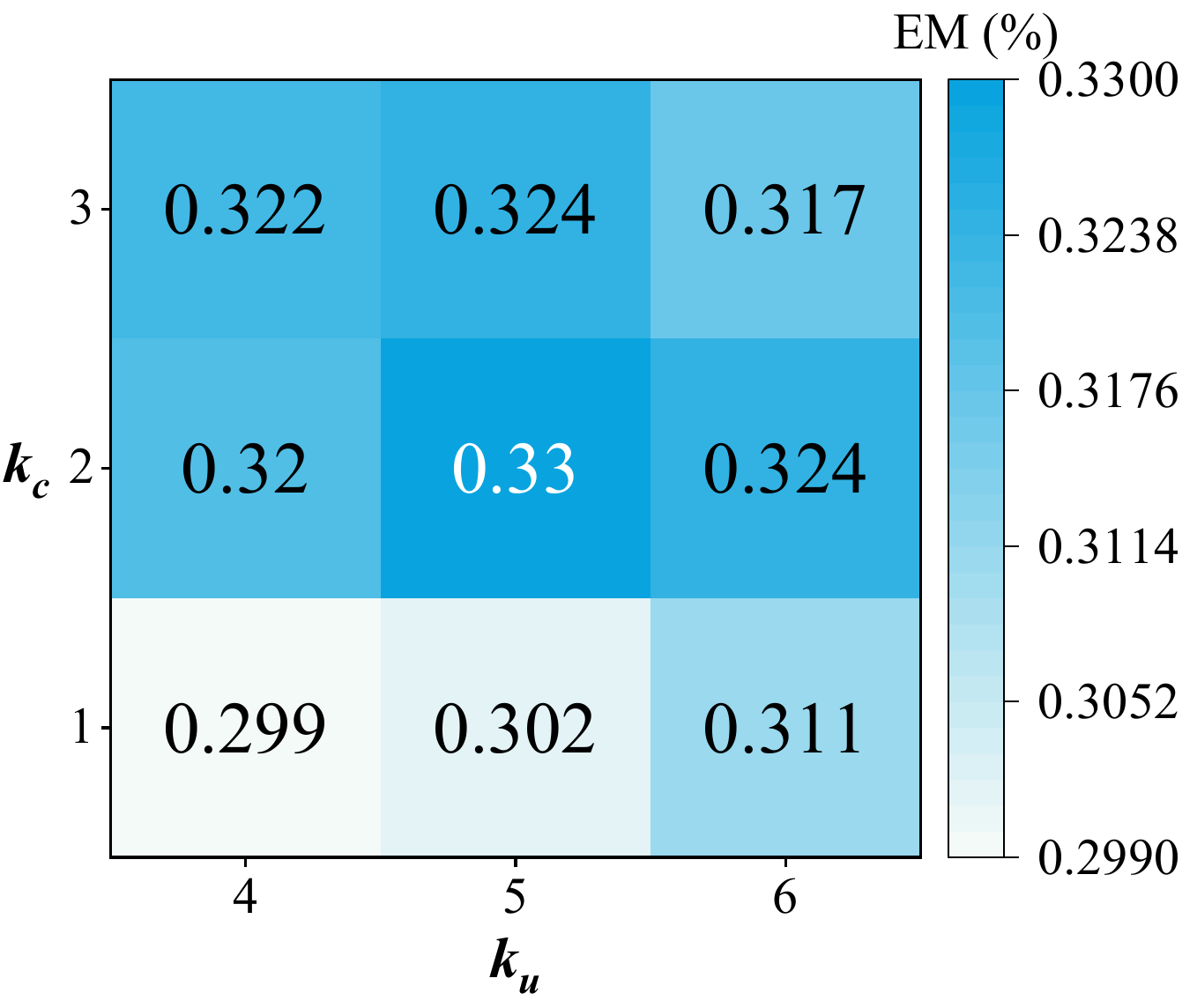}
        \caption{Amplification Coefficients}
    \end{center}
    \label{c_u_topk}
  \end{subfigure}\hfill
  \begin{subfigure}[b]{0.48\columnwidth}
    \begin{center}
        \includegraphics[width=0.92\linewidth]{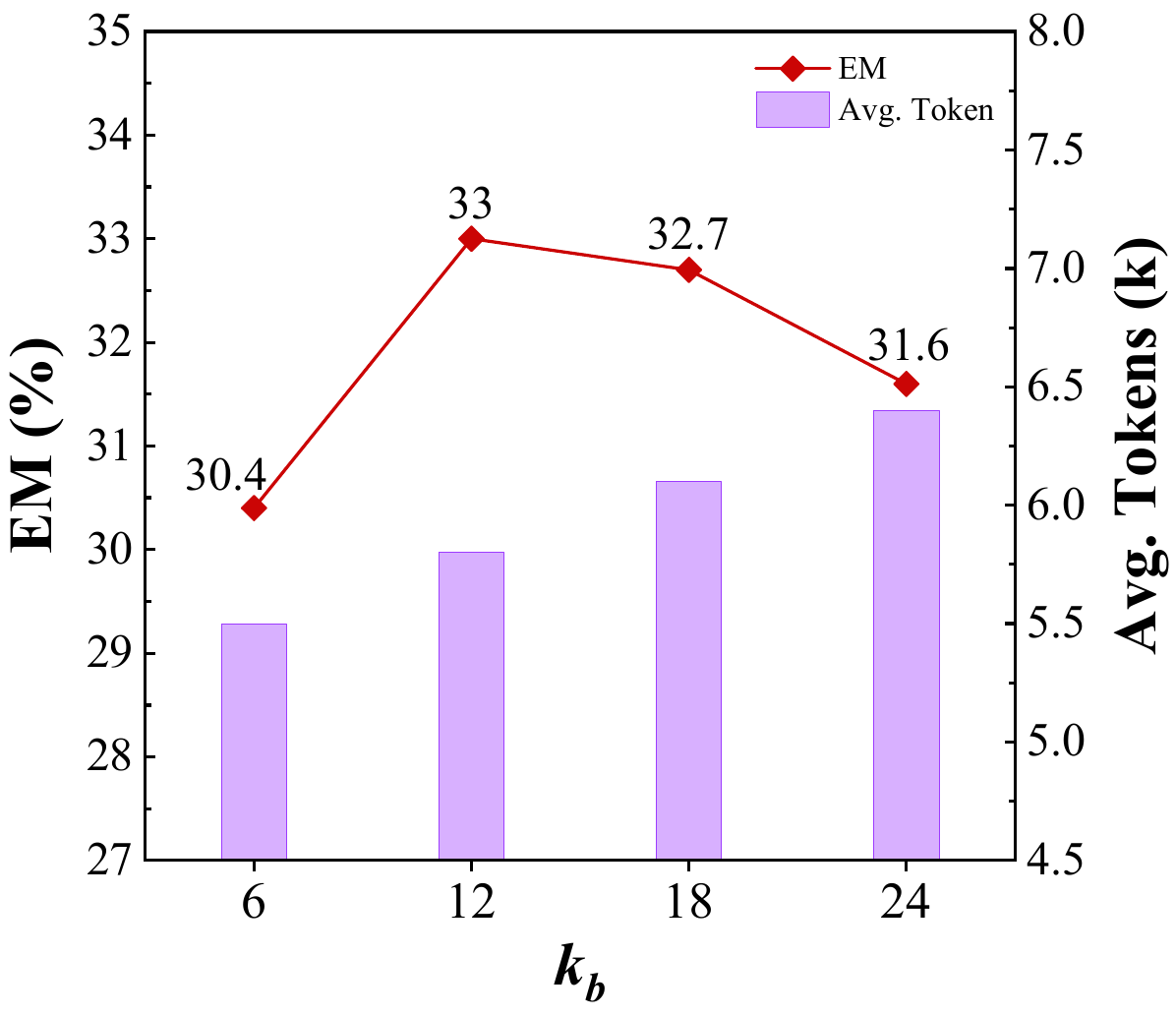}
        \caption{Base quota of DF-Retrieval}
    \end{center}
    \label{kbase}
  \end{subfigure}
  \caption{
    \textbf{Hyperparameter Sensitivity Analysis.} Subfigure\,($a$) is the heatmap for different combinations of $k_u$ and $k_c$, while subfigure\,($b$) shows the comparison results with different $k_b$.
  }
  \label{hyperparameter}
\end{center}
\end{figure}

\begin{figure}[ht]
  \centering
  \begin{center}
      \includegraphics[width=\columnwidth]{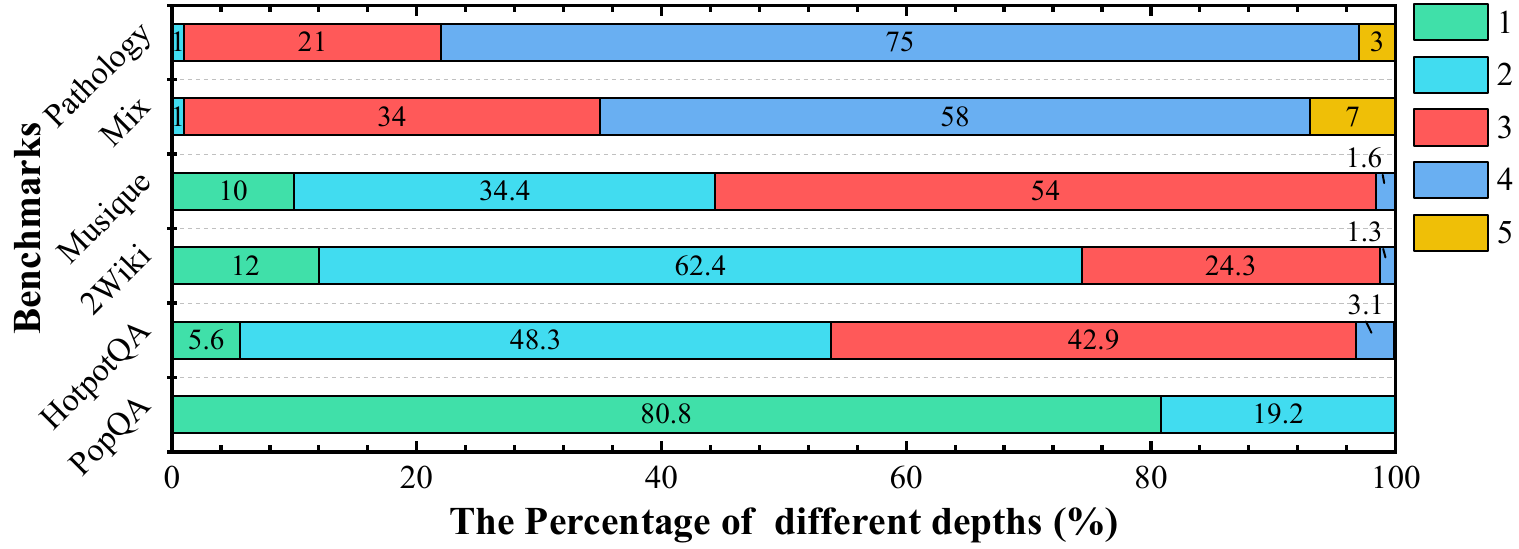}
      \caption{%
        The percentage of different depths on six benchmarks.
      }
      \label{depth_benchmarks}
  \end{center}
\end{figure}

\subsection{Controlling Amplification Coefficients}

While chunks link fragmented units to improve response completeness and fluency, oversized chunks risk overloading the LLM's semantic filter and burying critical evidence. Therefore, we set the amplification coefficients to $k_u=5$ and $k_c=2$, achieving the optimal trade-off between precision and readability as demonstrated in \cref{hyperparameter}\,(\subref{c_u_topk}). For DF-Retrieval, the quota $k_b$ influences both token costs and the correctness of the diffusion path. As shown in \cref{hyperparameter}\,(\subref{kbase}), performance peaks at the default setting of $k_b=12$.

\begin{figure}[ht]
  \centering
  \begin{subfigure}[b]{0.48\columnwidth}
    \begin{center}
        \includegraphics[width=\linewidth]{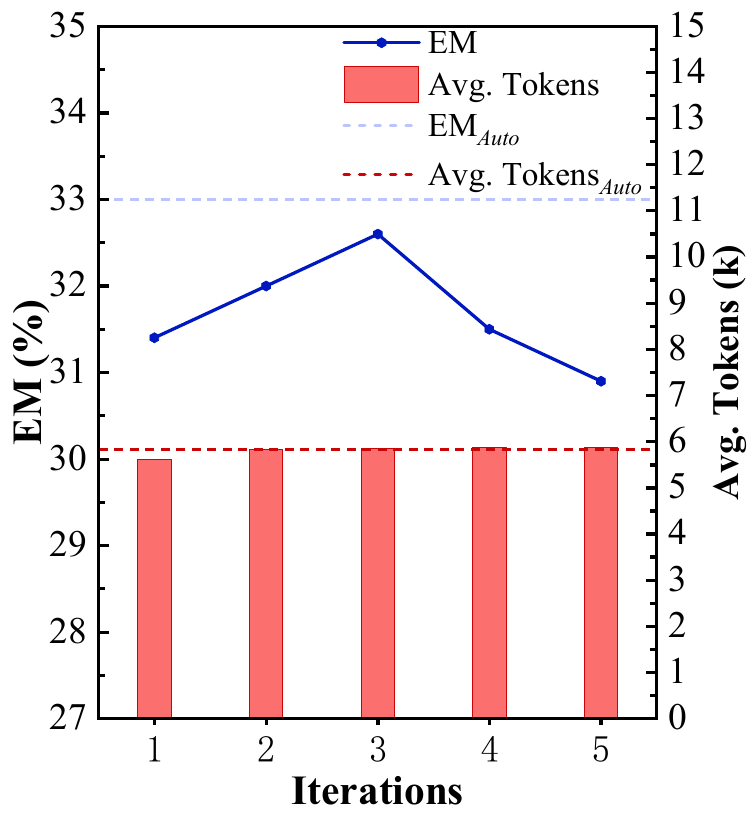}
        \caption{Control Iterations Only}
    \end{center}
    \label{iterations}
  \end{subfigure}\hfill
  \begin{subfigure}[b]{0.48\columnwidth}
    \begin{center}
        \includegraphics[width=\linewidth]{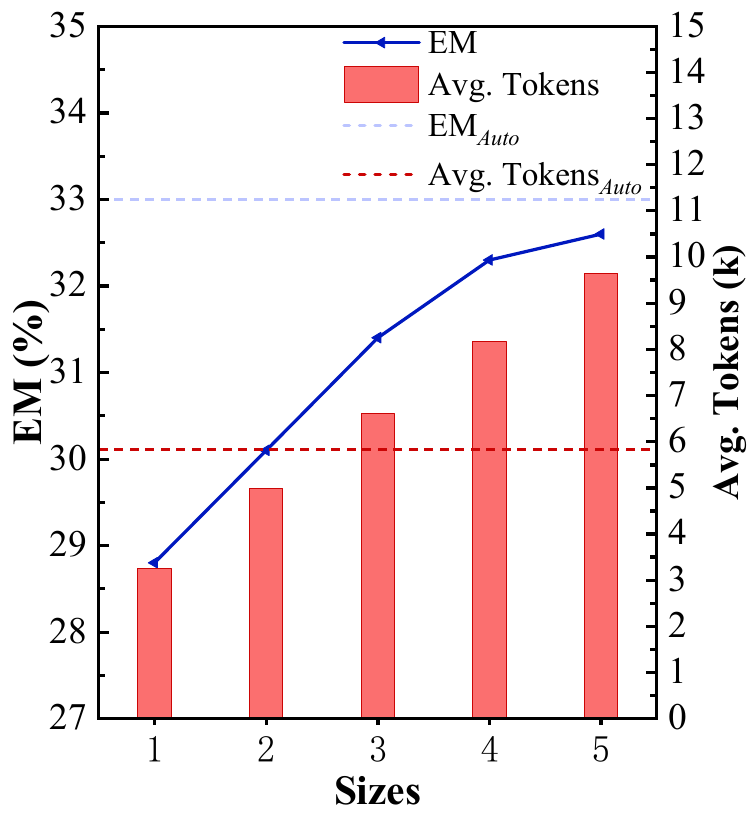}
        \caption{Control Sizes Only}
    \end{center}
    \label{sizes}
  \end{subfigure}
  \caption{
    \textbf{Ablation on Iterations and Window Sizes.} Impact of various semantic depth on QA performance and token costs.
  }
  \label{depth_comparison}
\end{figure}

\subsection{The Effectiveness of Adaptive Deep-mining}
%


To ascertain whether adaptive deep mining and dynamic context windows constitute effective, explainable mechanisms or compromise performance, we conducted controlled experiments on MuSiQue with a fixed retrieval strategy.

%


\cref{depth_comparison}\,(\subref{iterations}) shows that performance peaks at iteration $3$ and then declines. Although the fixed-iteration variant consumes a comparable number of tokens, it consistently lags behind IGMiRAG. The peak coincides with \cref{depth_benchmarks}, where IGMiRAG allocates the largest proportion of depth $3$ ($54.0\%$) on MuSiQue. This confirms that a fixed iteration budget cannot adapt to the true distribution of reasoning depths.


\cref{depth_comparison}\,(\subref{sizes}) presents that while performance initially improves with an expanding context window, the trend flattens. Even at the maximum token budget of $9.6\,\mathrm{k}^{+}$, the ceiling remains below that of IGMiRAG. 
These results confirm that marginal gains from naively stacking context degrade rapidly. In contrast, IGMiRAG’s complexity-driven window allocation achieves superior accuracy at a lower cost.

\section{Conclusion}

We propose IGMiRAG, a novel RAG framework designed to optimize reasoning depth and retrieval efficiency. By constructing a Hierarchical Heterogeneous Hypergraph to model human-like memory structures and employing an intuition-inspired strategy to guide associative diffusion, IGMiRAG introduces a ``Strategy-Diffusion'' paradigm. This paradigm enables in-depth memory mining with adaptive context scaling, achieving superior performance with fewer dynamic tokens. Such a cognitively inspired approach not only enhances retrieval efficiency and effectiveness but also bridges artificial mechanisms with human-like memory processing. Despite the approximate nature of current intuition signals, refining intuition-retrieval alignment holds promise for advancing LLM memory recall and complex reasoning.


\section*{Impact Statement}

This paper presents work on Retrieval-Augmented Generation (RAG), aiming to advance the field by improving retrieval efficiency and effectiveness to enhance memory precision and reasoning depth in Large Language Models. While our work may have various potential societal implications, we do not foresee specific concerns that warrant emphasis beyond the general risks associated with large language models and information retrieval systems.



\bibliography{example_paper}
\bibliographystyle{icml2026}

\newpage
\appendix
\onecolumn


\section*{Appendix}
Within this supplementary material, we elaborate on the following aspects:
\begin{itemize}
    \item Appendix \ref{PABD_Algorithm}: PABD Algorithm
    \item Appendix \ref{Pipeline_Example}: IGMiRAG Pipeline Example
    \item Appendix \ref{Case_Study}: Case Studies
    \item Appendix \ref{Experiment_Details}: Experimental Details
    \item Appendix \ref{Additional_Discussions}: Additional Discussions
    \item Appendix \ref{LLM_Prompts}: All LLM Prompts
\end{itemize}


\section{PABD Algorithm}
\label{PABD_Algorithm}

\begin{figure}[ht]
  \centering
  \begin{center}
      \includegraphics[width=\textwidth]{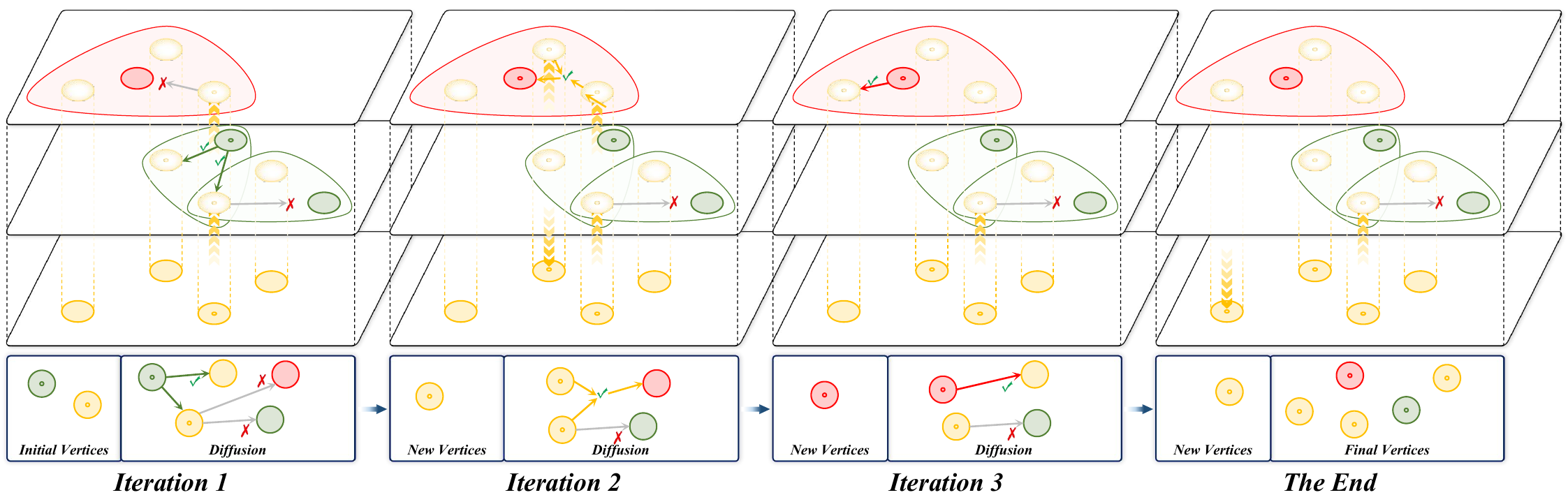}
      \caption{%
        An example workflow of the Preference-Aware Bidirectional Diffusion Algorithm.
      }
      \label{PABD_workflow}
  \end{center}
\end{figure}

To clearly elucidate the internal mechanisms of the PABD algorithm, we present a complete workflow example in \cref{PABD_workflow}, and provide corresponding pseudocode in \cref{alg_PABD}.
The PABD algorithm follows an alternating inference rule of ``top-down broadcasting and bottom-up screening''. Starting from the memory anchors, it performs bidirectional diffusion across knowledge levels at each iteration, enabling the deductive propagation of relevance scores.


Specifically, the score propagation within each iteration comprises two directed phases:
(i)The top-down diffusion process will broadcast scores from high-order anchors to their associated low-order knowledge; 
(ii)The bottom-up diffusion process will propagate from low-order anchors to high-order knowledge units that satisfy the threshold criteria.
The intensity of this propagation is governed by both a decay factor $\gamma$ and a preference coefficient $\rho$. While $\gamma$ is a fixed hyperparameter, $\rho$ is dynamically determined by the relevance quality of the target vertex's adjacent vertices (detailed in \cref{alg_pag}).

The bottom-up propagation is constrained by a dynamic threshold (illustrated in \cref{alg_dtf}). This threshold adaptively adjusts its bias based on activation feedback from the diffusion process: the bias increases upon activating new vertices to suppress the introduction of noise, and decreases otherwise to inject diffusion power. The specific bias regulation mechanism is detailed in \cref{alg_dtf}.

By combining this dynamic threshold mechanism with the preference-aware mechanism, the PABD algorithm effectively amplifies the scores of relevant paths while suppressing those of irrelevant ones, thereby revealing latent vertices.

Furthermore, upon the completion of each iteration, all newly activated vertices are merged with the current anchors to form the starting anchors for the subsequent iteration. The PABD algorithm terminates upon complete cessation of diffusion or upon reaching the maximum iterations, returning all activated vertices and their corresponding relevance scores in descending order.


\FloatBarrier  
\begin{algorithm*}[!htbp]
  \caption{Preference-Aware Bidirectional Diffusion (\textsc{PABD})}
  \label{alg_PABD}
  \begin{algorithmic}[1]
    \footnotesize
    \State {\bfseries Input:} initial scores $s(v)$,\; target layer $l$,\; max iteration $d$
    \State {\bfseries Output:} extended scores $s'(v)$

    \State {\bfseries Require:}
    \Statex \quad $\mathcal{G}_{\mathit{HHHG}}$\Comment{The Hierarchical Heterogeneous Hypergraph}
    \Statex \quad $id2l(\cdot)$ \Comment{The mapping from key to layer (all vertices)}
    \State {\bfseries Definitions:}
    \Statex \quad $\textsc{LFN}(u,\mathit{dir})$:\; layer-filtered neighbors of $u$ along $\mathit{dir}\in\{\textsc{forward},\textsc{backward}\}$
    \Statex \quad $\textsc{HC}(u,\mathit{dir})$:\; number of neighbors in $\textsc{LFN}(u,\mathit{dir})$ with score $> S[u]$
    \Statex \quad $\textsc{PA}(n)$:\; preference coefficient $\in[0,1]$, rising with $n$
    \Statex \quad $\textsc{PAG}(c,v,S,\gamma,\mathit{dir})$:\; score increment for forward/backward diffusion\Comment{Detailed in \cref{alg_pag}}
    \Statex \quad $\textsc{DTF}(c,\tau_0,l,b)$:\; returns {\sc True} if $c$ passes threshold $\tau$ under offset $b$\Comment{Detailed in \cref{alg_dtf}}
    \Statex \quad $\textsc{DTA}(\mathcal{V}_{\!F},\mathcal{V}_{\!B},\mathcal{V}_{\!A},l2id,b,i,\mathit{dir})$:\; updates $\mathcal{V}_{\!A},l2id,b,i$ with newly activated nodes\Comment{Detailed in \cref{alg_dta}}

    \State {\bfseries Initialize:}
    \Statex \quad $S \gets s(v)$;\; $\mathcal{V}_{\!A} \gets \mathrm{keys}(S)$
    \Statex \quad build $l2id[\ell],\;\ell\in\{1,2,3\}$ from $\mathcal{V}_{\!A}$ via $id2l(\cdot)$\Comment{The mapping from layer to key (activated vertices)}
    \Statex \quad decay factor $\gamma \gets 0.2$
    \Statex \quad thresholds $\tau^{(L)}\gets 0.5$,\; $\tau^{(H)}\gets 0.4$
    \Statex \quad bias $b \gets 0$

    \For{$i = 0$ {\bfseries to} $d$}
        \State $S' \gets S$;\; $\mathcal{V}_{\!F}\gets\emptyset$;\; $\mathcal{V}_{\!B}\gets\emptyset$

        \State {\bfseries Top-Down: Forward Diffusion Stage}\Comment{$\mathcal{L},\mathcal{H}\to\mathcal{N}$}
        \State $\mathcal{V}_{\!H}\gets l2id[3]\cup l2id[2]$\Comment{Get the high-order anchors}
        \For{$c\in\mathcal{V}_{\!H}$}
            \For{$v\in \Call{LFN}{c,\textsc{forward}}$}
                \If{$S[c]>S[v]$ {\bfseries and} $\Call{HC}{v,\textsc{forward}}>0$}
                    \State $S'[v] \gets S'[v] + \Call{PAG}{c,v,S,\gamma,\textsc{forward}}$\Comment{Update the scores of lower-level vertices}
                    \If{$v\notin \mathcal{V}_{\!A}$}
                        \State add $v$ to $\mathcal{V}_{\!F}$
                    \EndIf
                \EndIf
            \EndFor
        \EndFor
        \State $\Call{DTA}{\mathcal{V}_{\!F},\emptyset,\mathcal{V}_{\!A},l2id,b,\_,\textsc{forward}}$\Comment{Update the activated vertices, mapping, and bias}

        \State {\bfseries Bottom-Up: Backward Diffusion Stage}\Comment{$\mathcal{N}\to\mathcal{L},\mathcal{H}$}
        \State $\mathcal{V}_{\!L}\gets l2id[1]$\Comment{Get the low-order anchors}
        \For{$v\in\mathcal{V}_{\!L}$}
            \For{$c\in \Call{LFN}{v,\textsc{backward}}$}
                \If{$S[v]>S[c]$ {\bfseries and} $\Call{DTF}{c,\tau_0,l,b}$}
                    \State $S'[c] \gets S'[c] + \Call{PAG}{c,v,S,\gamma,\textsc{backward}}$\Comment{Update the scores of higher-level vertices}
                    \If{$c\notin \mathcal{V}_{\!A}$}
                        \State add $c$ to $\mathcal{V}_{\!B}$
                    \EndIf
                \EndIf
            \EndFor
        \EndFor
        \State $\Call{DTA}{\emptyset,\mathcal{V}_{\!B},\mathcal{V}_{\!A},l2id,b,i,\textsc{backward}}$\Comment{Update the activated vertices, mapping, bias, and $i$}

        \State $S \gets S'$\Comment{Update the memory anchors for next iteration}

        \If{$\mathcal{V}_{\!F}=\emptyset$ {\bfseries and} $\mathcal{V}_{\!B}=\emptyset$ {\bfseries and} $b=0.50$}\Comment{Whether the termination condition has been met}

            \State {\bfseries break}
        \EndIf
        \State $i \gets i+1$\Comment{Next iteration}

    \EndFor
    \State $s'(v) \gets S$

    \State {\bfseries return} $s'(\cdot)$
  \end{algorithmic}
\end{algorithm*}


\begin{algorithm*}[!htbp]
  \caption{Preference-Aware Gain (\textsc{PAG})}
  \label{alg_pag}
  \begin{algorithmic}[1]
    \Procedure{PAG}{$c, v, S, \gamma, \mathit{dir}$}
      \If{$\mathit{dir} = \textsc{forward}$}\Comment{Forward}
          \State $n \gets \Call{HC}{v,\,\textsc{forward}}$ \Comment{HC is defined in \cref{alg_PABD}}
          \State $\rho \gets \Call{PA}{n}$\Comment{PA is defined in \cref{alg_PABD}}
          \State $s' \gets (S[c]-S[v]) \cdot \rho \cdot \gamma$
      \Else  \Comment{Backward}
          \State $n \gets \Call{HC}{c,\,\textsc{backward}}$
          \State $\rho \gets \Call{PA}{n}$
          \State $\mathcal{P} \gets n/\Call{LFN}{v,\,\textsc{backward}}$\Comment{Get the adjacent activation proportion}
          \State $s' \gets (S[v]-S[c]) \cdot (\rho \cdot 0.5 + \mathcal{P}^{2} \cdot 0.5) \cdot \gamma$
      \EndIf
      \State {\bfseries return} $s'$
    \EndProcedure
  \end{algorithmic}
\end{algorithm*}


\begin{algorithm*}[!htbp]
  \caption{Dynamic Threshold Filtering (\textsc{DTF})}
  \label{alg_dtf}
  \begin{algorithmic}[1]
    \Procedure{DTF}{$u$,\;$\tau_0$,\;$l$,\;$b$}
      \If{$l_u = l$}                      \Comment{$l_u$ is the layer of $u$}
          \State $b \gets b + 0.05$
      \EndIf
      \State $\tau' \gets \tau_0^{\,\mathit{t}} - b$  \Comment{$\mathit{t}$ is the type of $u$}
      \State $\textit{if\_Pass} \gets (\mathcal{P} > \tau')$
      \State {\bfseries return} $\textit{if\_Pass}$
    \EndProcedure
  \end{algorithmic}
\end{algorithm*}


\begin{algorithm*}[!htbp]
  \caption{Dynamic Threshold Adjusting (\textsc{DTA})}
  \label{alg_dta}
  \begin{algorithmic}[1]
    \State {\bfseries Definitions:}
        \Statex \quad $\textsc{UFA}(\mathcal{V}_{\!F},\mathcal{V}_{\!A},l2id)$:\; update $\mathcal{V}_{\!A}$ and $l2id$ with newly activated vertices $\mathcal{V}_{\!F}$
        \Statex \quad $\textsc{UBA}(\mathcal{V}_{\!B},\mathcal{V}_{\!A},l2id)$:\; update $\mathcal{V}_{\!A}$ and $l2id$ with newly activated vertices $\mathcal{V}_{\!B}$
    \Procedure{DTA}{$\mathcal{V}_{\!F}$,
                   $\mathcal{V}_{\!B}$,
                   $\mathcal{V}_{\!A}$,
                   $l2id$,
                   $b$,
                   $i$,
                   $\mathit{dir}$}
      \If{$\mathit{dir} = \textsc{forward}$}\Comment{Forward}
          \If{$\mathcal{V}_{\!F} \neq \emptyset$}
              \State $\Call{UFA}{\mathcal{V}_{\!F},\,\mathcal{V}_{\!A},\,
                                                    l2id)}$
          \Else
              \State $b \gets \min(b + 0.1,\; 0.5)$
          \EndIf
      \Else  \Comment{\textsc{backward}}
          \If{$\mathcal{V}_{\!B} \neq \emptyset$}
              \State $\Call{UBA}{\mathcal{V}_{\!B},\,
                                                     \mathcal{V}_{\!A},\,
                                                     l2id}$
              \State $b \gets \max(b - 0.10,\; 0.0)$
          \Else
              \State $b \gets \min(b + 0.15,\; 0.5)$
              \State $i \gets i - 1$\Comment{Backtracking}
          \EndIf
      \EndIf
      \State {\bfseries return} $b$, $i$\Comment{Return new $b$ and $i$}
    \EndProcedure
  \end{algorithmic}
\end{algorithm*}

\FloatBarrier  

\section{IGMiRAG Pipeline Example}
\label{Pipeline_Example}
\cref{index_example} and \cref{query_example} illustrate a representative example of the IGMiRAG indexing process and querying process, respectively.
In the querying process example, content directly related to the final answer is highlighted in green.

\begin{figure}[!htbp]   
  \begin{center}
    \begin{infopanel}{green!50!blue!99}{cog}{Indexing: Extracting knowledge from the original chunks}
    \textbf{Title: Portrait of George Dyer Talking}\newline
    \textbf{Text: }Portrait of George Dyer Talking is an oil painting by Francis Bacon executed in 1966. It is a portrait of his lover George Dyer made at the height of Bacon's creative power. It depicts Dyer sitting on a revolving office stool in a luridly coloured room. His body and face are contorted, and his legs are tightly crossed. His head appears to be framed within a window or door. Above him is a naked hanging lightbulb, a favourite motif of Bacon's. The work contains a number of spatial ambiguities, not least that Dyer's body seems to be positioned both in the fore- and background.
    \newline
    
    \textbf{-*Entities*--}
    \begin{adjustwidth}{0em}{0em}
    \begin{itemize}
      \item George Dyer
      \item Francis Bacon
    \end{itemize}
    \end{adjustwidth}
    
    \textbf{-*Pairwise Relations*-}
    \begin{adjustwidth}{0em}{0em}
    \begin{itemize}
      \item $<$George Dyer, Francis Bacon$>$
      \item $<$Francis Bacon, Portrait of George Dyer Talking$>$
    \end{itemize}
    \end{adjustwidth}
    \tcbline
    \textbf{Title: Francis Bacon}\newline
    \textbf{Text: }Francis Bacon was born on 22 January 1561 at York House near the Strand in London, the son of Sir Nicholas Bacon (Lord Keeper of the Great Seal) by his second wife, Anne (Cooke) Bacon, the daughter of the noted humanist Anthony Cooke. His mother's sister was married to William Cecil, 1st Baron Burghley, making Burghley Bacon's uncle.
    \newline
    
    \textbf{-*Entities*-}
    \begin{adjustwidth}{0em}{0em}
    \begin{itemize}
      \item Francis Bacon
      \item Anne (Cooke) Bacon
      \item Sir Nicholas Bacon
      \item William Cecil, 1st Baron Burghley
    \end{itemize}
    \end{adjustwidth}
    
    \textbf{-*Pairwise Relations*-}
    \begin{adjustwidth}{0em}{0em}
    \begin{itemize}
      \item $<$Francis Bacon, Sir Nicholas Bacon$>$
      \item $<$Francis Bacon, Anne (Cooke) Bacon$>$
      \item $<$Sir Nicholas Bacon, Francis Bacon$>$
      \item $<$Francis Bacon, William Cecil, 1st Baron Burghley$>$
    \end{itemize}
    \end{adjustwidth}
    
    \textbf{-*Multiple Associations*-}
    \begin{adjustwidth}{0em}{0em}
    \begin{itemize}
      \item $<$Francis Bacon, Sir Nicholas Bacon, Anne (Cooke) Bacon$>$
    \end{itemize}
    \end{adjustwidth}
    \tcbline
    \textbf{Title: Head I}\newline
    \textbf{Text:} Head I is a relatively small oil and tempera on hardboard painting by the Irish-born British figurative artist Francis Bacon. Completed in 1948, it is the first in a series of six heads, the remainder of which were painted the following year in preparation for a November 1949 exhibition at the Hanover Gallery in London. Like the others in the series, it shows a screaming figure alone in a room, and focuses on the open mouth. The work shows a skull which has disintegrated on itself and is largely a formless blob of flesh. The entire upper half has disappeared, leaving only the jaw, mouth and teeth and one ear still intact...
    \newline
    
    \textbf{-*Entities*-}
    \begin{adjustwidth}{0em}{0em}
    \begin{itemize}
      \item Head I
      \item Francis Bacon
    \end{itemize}
    \end{adjustwidth}
    
    \textbf{-*Pairwise Relations*-}
    \begin{adjustwidth}{0em}{0em}
    \begin{itemize}
      \item $<$Francis Bacon, Head I$>$
    \end{itemize}
    \end{adjustwidth}
    
    \end{infopanel}
  \caption{\textbf{An Example of the Indexing Process.}}
  \label{index_example}
  \end{center}
\end{figure}

\begin{figure}[!htbp]
  \begin{center}

    \begin{infopanel}{green!35!blue!99}{search}{Quering: Answering by retrieving knowledge (including original chunks)}
    \textbf{Question: Who was the father of The Portrait of George Dyer Talking's creator?}
    \tcbline
    \textbf{rewrite\_question:} Who is the father of Francis Bacon, the creator 
    of the artwork titled 'The Portrait of George Dyer Talking'?
    \newline
    \textbf{key\_entities:} Francis Bacon \textbar Father, The Portrait of George Dyer Talking \textbar Artwork \textbar Painting
    \newline
    \textbf{keywords:} father, Francis Bacon, The Portrait of George Dyer Talking, 
    creator, art
    \newline
    \textbf{target\_layer:} 1,\quad
    \textbf{matching\_score:} 4,\quad
    \textbf{semantic\_depth:} 2
    \tcbline
    \textbf{-*Entities*--}
    \begin{adjustwidth}{0em}{0em}
    \begin{itemize}
      \item George Dyer
      \item \mybox[green]{Francis Bacon}
      \item Anne (Cooke) Bacon
      \item Diana Watson
      \item Head I
      \item William Cecil, 1st Baron Burghley
      \item Albrecht Dürer
      \item Albrecht Dürer the Elder
      \item Dürer's Father
      \item Portrait of Dürer's Father at 70
      \item \mybox[green]{Sir Nicholas Bacon}
    \end{itemize}
    \end{adjustwidth}
    
    \textbf{-*Pairwise Relations*-}
    \begin{adjustwidth}{0em}{0em}
    \begin{itemize}
      \item $<$Francis Bacon, Portrait of George Dyer Talking$>$
      \item $<$George Dyer, Francis Bacon$>$
      \item $<$Francis Bacon, 1945 painting$>$
      \item $<$Francis Bacon, Anne (Cooke) Bacon$>$
      \item \mybox[green]{$<$Francis Bacon, Sir Nicholas Bacon$>$}
      \item $<$Francis Bacon, Head I$>$
      \item $<$George Dyer, Albrecht Dürer$>$
      \item $<$Albrecht Dürer, Albrecht Dürer the Elder$>$
    \end{itemize}
    \end{adjustwidth}
    
    \textbf{-*Multiple Associations*-}
    \begin{adjustwidth}{0em}{0em}
    \begin{itemize}
      \item \mybox[green]{$<$Francis Bacon, Sir Nicholas Bacon, Anne (Cooke) Bacon$>$}
      \item $<$Figure in a landscape, Francis Bacon, Eric Hall, Diana Watson$>$
    \end{itemize}
    \end{adjustwidth}
    
    \textbf{-*Passages*-}
    \begin{adjustwidth}{0em}{0em}
    \begin{itemize}
      \item Title: Portrait of George Dyer Talking
      \item \mybox[green]{Title: Francis Bacon}
      \item Title: Head I
      \item Title: Figure in a landscape
      \item Title: Portrait of Dürer's Father at 70
    \end{itemize}
    \end{adjustwidth}
    \tcbline
    \textbf{-*LLM Output*-}
    \begin{adjustwidth}{0em}{0em}
    \begin{itemize}
      \item Thought: The creator of ``Portrait of George Dyer Talking'' is Francis Bacon. To determine his father, I refer to the information about Francis Bacon's family lineage. According to the data, \mybox[green]{Francis Bacon was the son of Sir} \mybox[green]{Nicholas Bacon}, who held the position of Lord Keeper of the Great Seal. Therefore, Sir Nicholas Bacon is the father of Francis Bacon, the artist who created ``Portrait of George Dyer Talking.'' 
      \item Answer: Sir Nicholas Bacon.
    \end{itemize}
    \end{adjustwidth}
    
    \end{infopanel}

      \caption{\textbf{An Example of the Quering Process.} The highlighted green area contains information directly related to the answer.}
      \label{query_example}
  \end{center}
\end{figure}

\section{Case Study}
\label{Case_Study}

To illustrate the advantages of IGMiRAG, representative cases across three tasks are presented. Semantic elements in the figure are color-coded as follows: green regions denote the correct knowledge matching the standard answer; yellow regions indicate irrelevant or information-deficient contents; and red regions represent misleading noise fragments that misguide the model.

As demonstrated in \cref{simpleqa_case_study} and \cref{reasonqa_case_study}, IGMiRAG generates responses faithful to the standard answers by synthesizing logically coherent and highly relevant retrieval results. In contrast, NodeRAG fails to derive correct conclusions due to fragmented reasoning stemming from missing critical supporting facts. Furthermore, Hyper-RAG produces compromised outputs, as the injection of irrelevant noise leads to misjudgment.

In the Knowledge Explanation task (\cref{explanatoryqa_case_study}), the blue areas highlight critical information points essential for a complete answer. IGMiRAG's response precisely covers all key points without redundancy. In contrast, NodeRAG omits two core pieces of information—“Ritika's father's name” and “the film's screenwriter”—while introducing irrelevant descriptions and factual inaccuracies absent from the gold standard. More severely, Hyper-RAG fails to address the film's core plot (e.g., “the conflict between Prabhu and Rithika's father”), instead generating extensive abstract summaries. This deficiency stems from Hyper-RAG's diffusion process lacking relevance constraints, causing retrieved fragments to deviate significantly from the query. Consequently, the model is forced to produce vacuous, hallucinatory summaries driven by noise.

\begin{figure}[!htbp]
  \begin{center}
    \begin{infopanel}{red!65!blue!99}{book}{Case study on Detail Capture}
    \textbf{--Details Capture--}\newline
    \textbf{Question:} In what city was Ray Strauss born?\newline
    \textbf{Gold\_Answer:} [``Perth, Australia'', ``Perth, Western Australia'', ``Perth'', ``Perth, W.A.'']
    \begin{adjustwidth}{0em}{0em}
    \begin{itemize}
        \item \textbf{IGMiRAG:} Thought: The provided data mentions that \mybox[green]{Raymond Bernard Strauss was from Perth}, Western Australia. It describes his background and ties to the area where he played cricket and field hockey. Since Ray Strauss is identified with this individual and no other city is mentioned in association with his birth, I can deduce that he was born in the same city as listed. Answer: \mybox[green]{Perth.}
        \item \textbf{NodeRAG:} Thought: The query is about the birthplace of Ray Strauss. From the information provided, specifically referencing notable individuals named Strauss, it is mentioned that Ray Strauss was an Australian cricketer. While the specific city of his birth is not indicated directly in the information given, we can infer from his nationality that he was likely born in an Australian city. However, to provide an accurate answer, \mybox[yellow]{I would need more precise details on Ray Strauss's birthplace, which are not included in the data presented.} Since the data does not provide the exact city where Ray Strauss was born, I cannot conclude definitively. Answer: \mybox[yellow]{Unable to determine.}
        \item \textbf{HyperRAG:} Thought: The data does not explicitly state the city where Ray Strauss was born. However, it does mention that he was involved in local governance in Buckeye City, Arizona. It's essential to infer that while he has a notable connection to Buckeye City, the birthplace might not be directly stated in the available information. \mybox{Given that he represents Buckeye City, I will assume it is also where he was likely born} unless otherwise specified. However, I should note that the details regarding his birthplace are absent, as there are no specific entries related to his exact birthplace. Answer: \mybox{Buckeye City.}
    \end{itemize}
    \end{adjustwidth}
    
    \end{infopanel}
      \caption{\textbf{A Case Study on Detail Capture Task.}}
      \label{simpleqa_case_study}
  \end{center}
\end{figure}

\begin{figure}[!htb]
  \begin{center}
    \begin{infopanel}{red!50!blue!99}{book}{Case study on Multi-Hop Reasoning}
    \textbf{--Multi-Hop Reasoning--}\newline
    \textbf{Question:} Who is the chief judge?\newline
    \textbf{Gold\_Answer:} [``Honorable Justice Abiodun Smith'']
    \begin{adjustwidth}{0em}{0em}
    \begin{itemize}
        \item \textbf{IGMiRAG:} Thought: Friday Ossai Osanebi was \mybox[green]{born in Nigeria}, specifically representing Ndokwa East Local Government Constituency in \mybox[green]{the Delta State House of Assembly.} To determine the chief judge of his birthplace, I need to identify the current chief judge of Delta State. According to the data, Rosaline Patricia Irorefe Bozimo was the Chief Justice of Delta State until her retirement on January 1, 2011, and \mybox[green]{she was succeeded by Honorable Justice Abiodun Smith.} Therefore, the current chief judge of Delta State, which is the likely birthplace of Friday Osanebi, is Honorable Justice Abiodun Smith. Answer: \mybox[green]{Honorable} \mybox[green]{Justice Abiodun Smith.}
        \item \textbf{NodeRAG:} Thought: Friday Ossai Osanebi is \mybox[green]{a politician from Nigeria}, specifically representing Ndokwa East Local Government Constituency \mybox[green]{in Delta State.} His birthplace, although not explicitly mentioned, can be inferred to be in Delta State, Nigeria. As of 2016, the acting Chief Judge of Rivers State is Adama Lamikanra, succeeded by Daisy W. Okocha, \mybox[yellow]{but this information is not related to Delta State. There is no information} \mybox[yellow]{provided about the Chief Judge specifically for Delta State}, where Friday Osanebi is likely from. Therefore, since the directly related information regarding Chief Judges does not include Delta State or a specific individual, I cannot provide a definitive name based on the current data. Answer: Information \mybox[yellow]{not provided for Chief Judge} of Delta State.
        \item \textbf{HyperRAG:} Thought: Friday Ossai Osanebi is \mybox[green]{a politician from Nigeria}, and he was \mybox[green]{born in the Delta State} \mybox[green]{region}...Looking at the provided data, the entry for ``CHIEF JUSTICE OF DELTA STATE'' indicates the head of the judiciary in that jurisdiction, \mybox[red]{which previously mentioned was Daisy W. Okocha}, noted as the first female Chief Judge of Rivers State. However, to clarify, the term ``chief judge'' often refers to a similar position within the context of a state or court...\mybox[yellow]{If Daisy W. Okocha's term overlapssignificantly or continues dominantly} \mybox[yellow]{as recognized}, then it would typically provide context upon which current legal proceedings reflect. The assertion can be made that Daisy W. Okocha or someone explicitly listed in the database would fulfill the role, thus needing the reference for the title within Delta pursuits distinctly. Answer: \mybox[red]{Chief Judge of Delta State (Daisy W.} \mybox[red]{Okocha was significant historically).}

    \end{itemize}
    \end{adjustwidth}
    \end{infopanel}
      \caption{\textbf{A Case Study on Multi-Hop Reasoning Task.}}
      \label{reasonqa_case_study}
  \end{center}
\end{figure}

\begin{figure}[!htbp]
  \begin{center}
    \begin{infopanel}{red!35!blue!99}{book}{Case study on Knowledge Explanation}
    \textbf{--Knowledge Explanation--}\newline
    \textbf{Question:} What is the significance of Manimaran's directorial debut Udhayam NH4 in understanding the themes of youthful rebellion and familial conflict in Tamil cinema, and how does the character development of Prabhu reflect these themes during the narrative?
    \newline
    \textbf{Gold\_Answer:} Udhayam NH4 marks \mybox[blue]{the directorial debut of Manimaran in the Tamil film industry}, recognized for its narrative that centers \mybox[blue]{on youthful rebellion and the pursuit of love amidst adversity}, reflecting the changes in the protagonist's character throughout the storyline. Specifically, after encountering various conflicts, \mybox[blue]{Prabhu and} \mybox[blue]{Rithika's relationship} evolves significantly, showing not only their challenges stemming from familial disapproval but also highlighting their determination to fight for their love despite societal limitations. The film, produced under \mybox[blue]{the banner of Meeka Entertainment and Grass Root Film Company}, demonstrates the struggle against traditional family dynamics, which manifests through palpable tensions between \mybox[blue]{Prabhu and Rithika's father, Avinash Gowda}, further complicated by \mybox[blue]{an aggressive law enforcement character, ACP Manoj Menon.}
    \begin{adjustwidth}{0em}{0em}
    \begin{itemize}
        \item \textbf{IGMiRAG:} \mybox[green]{Manimaran's directorial debut}, Udhayam NH4, serves as a significant entry point into the themes of \mybox[green]{youthful rebellion and familial conflict within Tamil cinema.} The film revolves around \mybox[green]{Prabhu...who becomes} \mybox[green]{ enmeshedin a turbulent romantic relationship with his classmate Rithika} amid opposition from \mybox[green]{her father,} \mybox[green]{Avinash Gowda}...Prabhu’s character embodies the struggle of modern youth navigating love and the pressures imposed by tradition and authority figures...His relentless pursuit to elope with Rithika...underscores the psychological turmoil faced by young individuals when confronted with societal norms...Prabhu’s development—from a lovesick youth to a determined individual willing to risk everything for love—reflects the internal struggles of many young people...the film deftly intertwines these themes with thrilling elements, demonstrated through the involvement of \mybox[green]{law enforcement, particularly ACP Manoj Menon}...\mybox[green]{Manimaran, with the collaboration of} \mybox[green]{Vetrimaaran—who} wrote the film's script and screenplay—successfully crafts a narrative that resonates with the youth...Udhayam NH4 effectively articulates the struggles between personal passion and societal duty...Prabhu’s character arc serves as a poignant reflection of modern-day challenges faced by young people...

        \item \textbf{NodeRAG:} \mybox[green]{Manimaran's directorial debut}, ``Udhayam NH4,'' serves as a pivotal representation of \mybox[green]{youthful} \mybox[green]{rebellion and familial conflict within Tamil cinema.} The film centers on Prabhu, a character who embodies the struggles faced by young individuals caught between their desires and familial expectations. \mybox[green]{Prabhu's love} \mybox[green]{for Rithika} is initially thwarted by \mybox[yellow]{her father's disapproval}, encapsulating the intense familial conflict...As the narrative unfolds, \mybox[green]{Prabhu's determination to elope with Rithika} signifies not only his personal fight for love but also a challenge to conventional upbringing and family expectations...\mybox[yellow]{the authoritative figure of Manoj}, the encounter specialist assigned to contain their budding romance...\mybox[yellow]{Each confrontation between Manoj and Prabhu} underscores the film's thematic exploration of rebellion as \mybox[red]{Prabhu strives to take control of his fate}...Overall, ``Udhayam NH4'' utilizes Prabhu's character development to delve into the complexities of growing up, where familial conflict often fuels rebellion...Manimaran’s narrative thus becomes a lens through which the audience can explore the intricacies of youthful rebellion and the weight of familial obligations...

        \item \textbf{HyperRAG:} \mybox[green]{Manimaran's directorial debut}, ``Udhayam NH4,'' holds significant relevance in exploring the themes of \mybox[green]{youthful rebellion and familial conflict}...The narrative...portrays the journey of Prabhu, a young man entangled in a conflict driven by love and familial loyalties...Prabhu's character encapsulates the struggle between personal desires and family expectations...\mybox[green]{Prabhu and his love interest, Rithika}, decide to elope...marking their defiance against traditional familial structures...Prabhu’s character development illustrates a poignant evolution from a carefree youth towards a more conscious and resolute individual...Prabhu’s rebellious act of eloping signifies a larger commentary on the youth's quest for autonomy against societal norms...\mybox[green]{interactions} \mybox[green]{between Prabhu and...Manoj, a law enforcement officer} representing family authority, amplify the generational clash...Manoj's pursuit of Prabhu...symbolizes...the struggle against the limitations imposed by family expectations...In summary, ``Udhayam NH4'' is instrumental in enhancing the discourse on youthful rebellion and familial conflict...Through the character of Prabhu, Manimaran masterfully illustrates the complexities of navigating relationships...\mybox[red]{offering reflections that resonate with youths}...This thematic exploration marks \mybox[red]{``Udhayam NH4'' as a pivotal contribution to the evolving narrative landscape of Tamil cinema}...

    \end{itemize}
    \end{adjustwidth}
    \end{infopanel}
      \caption{\textbf{A Case Study on Knowledge Explanation Task.}}
      \label{explanatoryqa_case_study}
  \end{center}
\end{figure}

\section{Experiment Details}
\label{Experiment_Details}
\subsection{Baselines}
\label{Baselines_Details}
A comparative analysis was conducted against state-of-the-art and widely adopted RAG methods. Details regarding these baselines are provided below.

\textbf{NaiveRAG} \cite{lewis2020retrieval}: This baseline method segments the source text into fine-grained chunks. Each chunk is then embedded into a vector representation and stored in a vector database. During the retrieval phase, the system identifies the most similar vectors and concatenates the corresponding chunks to construct the final context.

\textbf{LightRAG} \cite{guo2024lightrag}: A standard graph-enhanced RAG method. It leverages graph structures to capture pairwise relationships among knowledge units, storing both entities and relations as vectors in two separate databases. The method employs a two-stage retrieval strategy: first, it retrieves entity and relation candidates based on vector similarity; next, it explores the graph starting from these candidates, gathering adjacent nodes and edges to form the final retrieved context.

\textbf{PathRAG} \cite{chen2025pathrag}: A graph-enhanced RAG method that employs a flow-control pruning mechanism.
It constructs its index identically to LightRAG. During retrieval, the system first identifies candidate entities and relations via keyword search. Subsequently, the flow-control pruning mechanism retains only the most relevant paths, with the final result comprising the sequence of entities and relations along these paths.

\textbf{NodeRAG} \cite{xu2025noderag}: A graph-enhanced RAG method that operates on heterogeneous graphs.
It employs such a graph structure to uniformly model all knowledge units and generates high-level community summaries through clustering. During retrieval, the system first aggregates global candidates using both keyword search and vector similarity. Subsequently, a shallow Personalised PageRank (PPR) is executed over these candidates to mine potential nodes, yielding the final retrieval context.

\textbf{Hyper-RAG} \cite{feng2025hyper}: A standard hypergraph-enhanced RAG method.
It models both pairwise and higher-order relationships as hyperedges, using an index construction process similar to LightRAG. During retrieval, the system first retrieves entity candidates and pairwise/multi-entity relation candidates separately via vector similarity. It then propagates activation through the hyperedges to generate the final set of entities and relations.

\textbf{Cog-RAG} \cite{hu2025cog}: A hypergraph-enhanced RAG method that employs dual hypergraph modeling for topics and entities. Building on Hyper-RAG, it leverages a topic hypergraph to capture topic associations within text chunks and adopts a two-stage retrieval process inspired by cognitive mechanisms. First, the system retrieves global candidates from the topic hypergraph; it then further retrieves fine-grained, relevant information from the entity hypergraph.

\subsection{Benchmarks}
\label{Benchmarks_Details}

We evaluated all methods on six benchmarks spanning three task types, with statistics details for each benchmark provided in \cref{statistics_tabel}.
For the Explanatory QA task, Mix comprises fragmented paragraphs sampled from 61 distinct domains, while Pathology is a domain-specific benchmark focused on pathology, containing only a single document.

\begin{table}[!htbp]
    \begin{center}
        \caption{
        The statistics details for each benchmark.
        }
        \label{statistics_tabel}

\begin{tabular}{lcccccc}

\toprule
\multirow{2}{*}{\textbf{Statistics}}
& \multicolumn{1}{c}{\textbf{Simple QA}}
& \multicolumn{3}{c}{\textbf{Multi-Hop QA}}
& \multicolumn{2}{c}{\textbf{Explanatory QA}} \\
\cmidrule(lr){2-2}\cmidrule(lr){3-5}\cmidrule(lr){6-7}
& PopQA & MuSiQue & 2Wiki & HotpotQA & Mix & Pathology \\
\midrule
\#~Passages (Documents)  & \num{8676} & \num{11656} & \num{6119} & \num{9811} & \num{61} & \num{1}\\
\#~Chunks    & \num{1780} & \num{2191} & \num{1056} & \num{2096} & \num{980} & \num{1099}\\
\#~Tokens    & \num{1184847} & \num{1362612} & \num{684059} & \num{1306249} & \num{615355} & \num{905760}\\
\#~Entities    & \num{41315} & \num{43827} & \num{22041} & \num{39899} & \num{18792} & \num{20556}\\
\#~Pairwise Relations    & \num{16697} & \num{21941} & \num{9843} & \num{19941} & \num{8820} & \num{10510}\\
\#~Multiple Associations    & \num{6654} & \num{8941} & \num{4359} & \num{7566} & \num{3373} & \num{4158}\\
\#~Queries   & \num{1000} & \num{1000} & \num{1000} & \num{1000} & \num{100} & \num{100}\\
\bottomrule



\end{tabular}
    \end{center}
\end{table}


\textbf{PopQA:} We randomly sampled \num{1000} queries following HippoRAG\,2 \cite{gutierrez2025rag}. This benchmark focuses on single-entity descriptive questions, designed to evaluate entity recognition and retrieval capabilities in the context of simple question answering.

\textbf{MuSiQue, 2WikiMultiHop and HotpotQA:} Following prior work \cite{press2023measuring,jimenez2024hipporag}, we randomly sampled \num{1000} queries from each validation set. These benchmarks primarily feature complex multi-hop questions that obscure target entities through multiple layers of relational embedding, thereby challenging both vector and keyword-based retrieval methods.

\textbf{Mix and Pathology:} We further refined the question generation methodology employed in Hyper-RAG and Cog-RAG. Consecutive, lengthy text segments were fed into an LLM, with instructions to generate 100 questions per reasoning hop (ranging from 1 to 3 hops) along with corresponding standard answers in the form of paragraph-level descriptions. We then randomly sampled \num{100} question-answer (QA) pairs to construct benchmarks with diverse hop counts. These benchmarks are designed to evaluate knowledge-explanation tasks that require extensive descriptive responses, necessitating models to integrate information across multiple contexts and produce coherent, detailed explanations. The prompt templates used for QA pair generation are provided in Appendix \ref{prompt_exp}.

\subsection{Metrics}
\label{Metrics_Details}
For detail-capture and multi-hop tasks, short-form outputs are evaluated using exact match (EM) and F1 scores computed directly against reference answers. In contrast, knowledge explanation tasks involve open-ended, long-form responses that cannot be reliably assessed via string matching. To address this, we adopt the LLM-as-a-judge paradigm: for each instance, the question, reference answer, and model-generated response are jointly fed into an LLM, which is prompted to score the output in terms of EM, precision, and recall—based on which F1 is then calculated. Detailed evaluation prompt templates are provided in Appendix \ref{prompt_exp}.
Across all benchmarks, we record the total token consumption of the model and normalize it by the number of questions to compute the average token cost per instance.

\subsection{Implementation Details and Hyperparameters}
\label{Implementation_Details}
For all baseline methods, we strictly follow the default hyperparameters and optimal operational modes specified in their official implementations. Detailed configurations are summarized in \cref{implementation_details}.

\begin{table}[!htbp]
    \begin{center}
        \caption{
        The implementation details of each baseline.
        }
        \label{implementation_details}

        \begin{tabular}{lcccccc}
        
        \toprule
        \textbf{Baselines} & Naive RAG & LightRAG & PathRAG & NodeRAG & Hyper-RAG & Cog-RAG \\
        \midrule
        Chunk Token Size  & \num{1200} & \num{1200} & \num{1200} & \num{1024} & \num{1200} & \num{1200}\\
        Chunk Overlap Token Size  & \num{100} & \num{100} & \num{100} & - & \num{100} & \num{100}\\
        \midrule
        Query Mode  & - & hybrid & hybrid & - & hyper & cog\\
        Top-k for Retrieve  & \num{1},\;\num{3},\;\num{5} & \num{60} & \num{40} & \num{10} & \num{60} & \num{60}\\
        Max Token for Local Context  & - & \num{1000} & \num{2500} & - & \num{300} & \num{300}\\
        Max Token for Global Context  & - & \num{2400} & \num{2100} & - & \num{1600} & \num{1600}\\
        Max Token for Text Context  & - & \num{2500} & \num{2100} & - & \num{1600} & \num{1600}\\
        \midrule
        Damping factor $\alpha$ of PPR & - & - & - & \num{0.5} & - & -\\
        Max Iterations of PPR & - & - & - & \num{2} & - & -\\
        \midrule
        LLM Model  & \multicolumn{6}{c}{\textbf{\textrm{GPT-4o-mini}}}\\
        Embedding Model  & \multicolumn{6}{c}{\textbf{\textrm{text-embedding-3-small}}}\\
        Temperature  & \multicolumn{6}{c}{\textbf{\textrm{0}}}\\
        \bottomrule
        
        
        
        \end{tabular}
        \end{center}
\end{table}

\section{Additional Discussions}
\label{Additional_Discussions}
Using the target layer $l$ and matching score $m$, DF-Retrieval dynamically allocates search quotas between global and target-layer local retrieval. To evaluate its effectiveness, we conducted comparative experiments with fixed $l \in [1,3]\cap \mathbb{Z}$ and $m \in [1, 5]\cap \mathbb{Z}$ on MuSiQue.

\begin{figure}[!htbp]
  \centering
  \begin{subfigure}[b]{0.48\columnwidth}
    \begin{center}
        \includegraphics[width=0.5\linewidth]{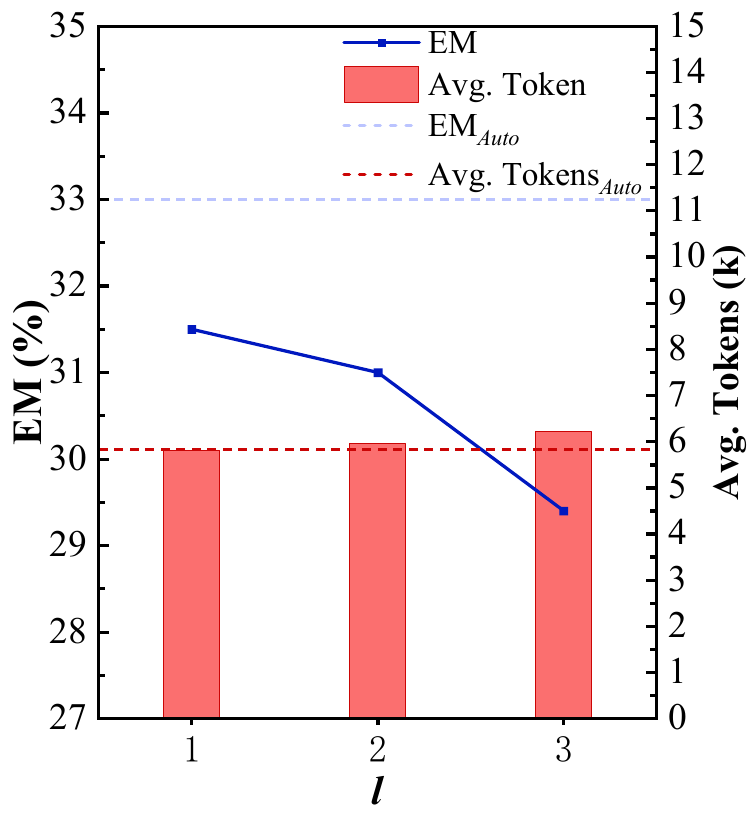}
        \caption{Control Target Layer Only}
    \end{center}
    \label{target_layer}
  \end{subfigure}\hfill
  \begin{subfigure}[b]{0.48\columnwidth}
    \begin{center}
        \includegraphics[width=0.5\linewidth]{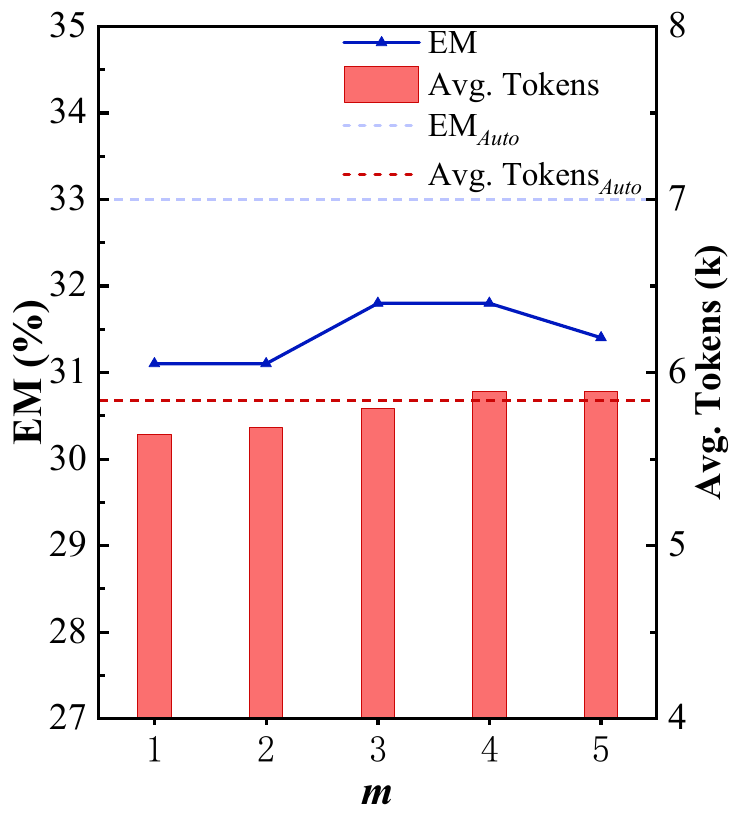}
        \caption{Control Matching Score Only}
    \end{center}
    \label{matching_score}
  \end{subfigure}
  \caption{
    \textbf{Ablation on target layer and matching score.} Impact of various target layer $l$ and matching score $m$ on QA performance and token costs.
  }
  \label{layer_mscore}
\end{figure}

\cref{layer_mscore}\,(\subref{target_layer}) illustrates the performance variation across different target layer settings. As the target layer increases, the Exact Match (EM) score exhibits a monotonically decreasing trend. This is attributed to the nature of MuSiQue queries, which involve multi-hop entity searches with nested, multi-layer relationships. Setting the target layer too high (e.g., at the binary relation or multi-event level) introduces misplaced focus and consequently increases noise, significantly degrading the quality of initial retrieval. These initial errors are then amplified during the subsequent diffusion stage, leading to a substantial performance decline.

As shown in \cref{layer_mscore}\,(\subref{matching_score}), any fixed quota configuration significantly underperforms IGMiRAG's dynamic strategy. Further analysis reveals that excessive bias toward either global or local granularity degrades retrieval effectiveness. These results indicate a strong complementarity between the two semantic spaces, suggesting that adaptively adjusting quotas based on task focus is crucial for enhancing retrieval quality.

\section{All LLM Prompts}
\label{LLM_Prompts}

\subsection{Prompts for Indexing}
\label{prompt_indexing}
\begin{myprompt_single}{Extracting Entities}
    \textbf{Prompt: }\textit{
    Identify as many entities as possible in the text, ensuring that no nouns with referential functions (including proper nouns, common nouns, place nouns, abstract concepts, etc.) are omitted.\newline
    \# \textbf{Special attention:}\newline
     - All pronouns must be forcibly parsed into the explicit entities they refer to, and converted into the form of ``specific reference + identity/role'' (such as: he $\to$ Tom'\,s father; this matter $\to$ the argument between Tom and Mickey);\newline
     - For implicit entities (such as objects indirectly referred to by verbs or adjectives, for example, ``complainer'', ``the criticized behavior''), if they play a semantic role in the context, they should also be extracted as independent entities;\newline
     - If a noun appears in a compound form (such as ``Beijing Municipal Government'', ``Grandma of Little Red Riding Hood''), it should be regarded as a whole entity and not be split;\newline
     - The description of the entity must include the entity name to ensure clarity.\newline
    For each entity, extract the following information:\newline
    \textbf{- entity\_name:} Use the complete expressions that appear in the original text. No translation or annotations are needed. If the name is long but has a clear referential meaning, retain the complete name. Choose the complete form for the first mention.\newline
    \textbf{- entity\_description:} Based on the content of the original text, describe the information of the entity in detail. The description should adopt the original sentence order and wording, provide a complete narrative without omitting any details; direct quotations of original phrases or appropriate supplements and analyses can be used, but the scattered information should be linked into a coherent, independent and readable paragraph. It is prohibited to summarize with a single sentence.\newline
    \textbf{- attribute:} List any additional information that is not mentioned in the description but may be important, such as time, location, status, category, etc. If nothing is available, then skip this item.
}
\end{myprompt_single}
\begin{myprompt_single}{Extracting Low-order Relations}
    \textbf{Prompt: }\textit{
    For all the entities determined in Step 1, conduct a full-scale pairwise analysis (that is, any two entities form a pair to be tested), and determine whether there are any grammatical or semantic relationships.\newline
\# \textbf{Special attention:}\newline
 - The types of relationships include but are not limited to:
 \begin{adjustwidth}{0em}{0em}
 \begin{itemize}  
  \item Grammatical relationships: subject–verb–object, copular–complement, existential, coordinate, comparative, etc., which are grammatical structure relationships involving two entities;
  \item Semantic relationships: synonymy, causality, sequence, containment, opposition, analogy, influence, etc., which are semantic structural associations involving two entities;
  \item Referential relationships: including explicit and implicit aliases.
\end{itemize}
\end{adjustwidth}
 - Different names of the same entity should also establish a pairing relationship! Including the following two types:
\begin{adjustwidth}{0em}{0em}
\begin{itemize}  
  \item Explicit Alias--Whenever parentheses, quotation marks, dashes, ``also known as'', ``abbreviation'', ``nickname'', or ``acronym'' are used, immediately capture them as synonymous entity pairs;
  \item Implicit Replacement--Observing the ``Reference Chain'': If the pronouns, names, or titles of the same subject change, it can be considered as a potential synonym.
\end{itemize}
\end{adjustwidth}
For each interrelated entity pair, extract the following information:\newline
\textbf{- entities\_pair:} The name of paired entities that must be consistent with the results obtained in Step 1. For example:``[entity1, entity2]''.\newline
\textbf{- relationship\_description:} Based on the analysis, describe the relationships between entities in detail. The description should adopt the original sentence order and wording to connect the two entities, provide a complete narrative without omitting any details; direct quotations of original phrases or appropriate supplements and analyses can be used, but the scattered information should be linked into a coherent, independent and readable paragraph. It is prohibited to summarize with a single sentence.\newline
\textbf{- attribute:} List any additional information that is not mentioned in the description but may be important, such as the time, place and context of the relationship. If the two entity names refer to the same object, then add the output content ``Same Entity'' in this item. If nothing is available, then skip this item.
}
\end{myprompt_single}
\begin{myprompt_single}{Extracting high-level Keywords}
    \textbf{Prompt: }\textit{
Based on all the identified entity relationships in Step 2, extract the high-level keywords that can summarize the main idea, core concepts, key events or abstract themes of the paragraph or the entire text.\newline 
\# Special attention:
\begin{adjustwidth}{0em}{0em}
\begin{itemize}  
  \item Keywords must be specific and have clear meanings, avoiding the use of vague terms such as ``emotion'', ``development'', or ``process'';
  \item By combining multiple relationship chains, theme-level concepts (such as ``family power struggle'', ``technology ethics dilemma'', ``urban-rural identity recognition'', etc.) can be extracted.
\end{itemize}
\end{adjustwidth}
\textbf{- high\_level\_keywords:} The content of high-level keywords should summarize the overall idea presented in the document. Avoid using specific names of certain entities that focus excessively on details, as well as vague or empty terms.
}
\end{myprompt_single}
\begin{myprompt_single}{Extracting High-order Relations}
    \textbf{Prompt: }\textit{
Based on the entities in Step 1, the entities pairs in Step 2 and the high-level keywords in Step 3, try to analyze their multi-dimensional connections or commonalities, and construct a high-order association set that includes as many entities as possible.\newline 
\# Association Types (satisfy one is sufficient):
\begin{adjustwidth}{0em}{0em}
\begin{itemize}  
  \item Event-driven Association: Multiple entities form a high-order association sequence connected together to form a semantic chain, including co-occurring events, event chains, multi-role events, and event nesting, etc.;
  \item Knowledge-logical Association: Multiple entities jointly constitute a certain knowledge system, theoretical framework, or logical structure, including concept composition, explanation of principles, classification system, and axiom system, etc.;
  \item Functional-cooperative Association: Multiple entities undertake different functions in a certain system, process, or organizational structure, collaborating to complete tasks, including process collaboration, system modules, organizational division of labor, and supply chain collaboration, etc.;
  \item Semantic Analogy \& Metaphor Association: Multiple entities form cross-domain mapping through analogy, metaphor, and symbolism, including structural analogy, metaphor mapping, and symbolic system, etc.;
  \item Stance \& Conflict Association: Multiple entities exhibit positions, interests, conflicts, or negotiation relationships on a certain issue, including position opposition, interest alliance, discourse contention, and role reversal, etc.;
  \item Temporal and Spatial Context Association: Multiple entities form non-event associations through common time, space, and cultural context, including simultaneous coexistence, cultural context, spatial topology, and time evolution, etc.;
  \item Other possible high-level associations among multiple entities that you think exist, as long as they are semantically coherent, are acceptable.
\end{itemize}
\end{adjustwidth}
\# Special attention:
\begin{adjustwidth}{0em}{0em}
\begin{itemize}  
  \item Each set must contain at least 4 entities, the more, the better;
  \item It is prohibited to forcibly connect entities without any intersection;
  \item Minimizing inter-group intersection: The intersection of the extracted high-order association sets should be as small as possible, and the redundancy should be minimized. It is not allowed for two sets to have an intersection that is nearly the same as the sets themselves.
\end{itemize}
\end{adjustwidth}
For each entities set, extract the following information:\newline 
\textbf{- entities\_set:} List every member entity exactly as named in Step 1.\newline 
\textbf{- relationship\_description:} Based on the relationships among the entities in the set, describe the relevant content represented by this set in detail, such as events, knowledge, functions, etc. The description should adopt the original sentence order and wording to connect all the entities within the set, and provide a complete narrative of at least 3 sentences without omitting any details; direct quotations of original phrases or appropriate supplements and analyses can be used, but the scattered information should be linked into a coherent, independent, and readable paragraph. It is prohibited to only list the relationships or summarize with a single sentence.\newline 
\textbf{- attribute:} List any additional information that is not mentioned in the description but may be important, such as the time, place, and background circumstances. If nothing is available, then skip this item.
}
\end{myprompt_single}

\subsection{Prompts for Strategy Parsing}
\label{prompt_parsing}
\begin{myprompt_single}{Knowledge Hypergraph Architecture}
    \textbf{Prompt: }\textit{
---Role---\newline 
You are a useful query analysis assistant, responsible for identifying the key entities, keywords, target layer and semantic depth of the user'\,s query problem based on the knowledge hypergraph architecture description provided to you.\newline 
---Knowledge Hypergraph Architecture---\newline 
 -*Hierarchical Hypergraph Architecture*-\newline 
  Layer 1 (Single Entity Layer)\newline 
   Answerable scope:
   \begin{adjustwidth}{0em}{0em}
       \begin{itemize}
           \item Entity identity, aliases, definition, category, attribute values, status, capabilities, constituent components.
           \item Key events in the entity'\,s life cycle, including ``time points'' and ``locations''.
           \item Version differences of the entity within the same work.
       \end{itemize}
   \end{adjustwidth}
  Layer 2 (Paired Entity Relationship Layer)
   Answerable scope:
   \begin{adjustwidth}{0em}{0em}
       \begin{itemize}
           \item Whether there is a specified relationship between two entities; relationship type, strength, direction, evidence sentences.
           \item Direct motives and results of a single interaction.
           \item Evolution of the relationship over time.
       \end{itemize}
   \end{adjustwidth}
  Layer 3 (Multi-Entity Event Layer)
   Answerable scope:
      \begin{adjustwidth}{0em}{0em}
       \begin{itemize}
           \item Event causality chain: background $\to$ trigger $\to$ process $\to$ result $\to$ impact (This event may not have been fully documented. The knowledge in this layer may only include a few parts of the causal chain, such as ``role $\to$ process $\to$ result'').
           \item Roles and conflicts of interest of each entity in the event.
           \item Quantitative impact of the event on subsequent plot or storyline.
       \end{itemize}
   \end{adjustwidth}
}
\end{myprompt_single}
\begin{myprompt_single}{Implicit Strategy Parsing}
    \textbf{Prompt: }\textit{
---Goal---\newline 
 -Initial query decomposition:\newline 
  \textbf{1.question:} Consistent with the original question.\newline 
  \textbf{2.rewrite\_question:} Without altering the original intention of the user, expand, clarify, structure and align the input query with entities, making it more convenient for downstream search engines or knowledge bases to retrieve, while enhancing semantic integrity.
  \begin{adjustwidth}{0em}{0em}
      \begin{itemize}
          \item Provide context (For example, time, place, topic, etc.).
          \item Entity alignment and period matching (If the entities in the query have specific names at different stages, then replace them with the original terms that match the corresponding stage).
          \item Disambiguation and Reference Clarification (Replace vague references such as ``it'', ``this'', and ``influence'' with specific nouns to avoid ambiguity.).
          \item Synonyms / Domain Terminology Expansion (Introduce synonyms, near-synonyms, and domain-specific terms to enhance the recall rate).
      \end{itemize}
  \end{adjustwidth}
  \textbf{3.key\_entities:} Extract nouns or noun phrases that have clear referential meanings from the query, such as people, places, organizations, events, technologies, concepts, etc.
  \begin{adjustwidth}{0em}{0em}
      \begin{itemize}
          \item The entity should be linkable (able to correspond to the entries in the knowledge base).
          \item Provide all the aliases, abbreviations, names, titles, etc. of this entity at different stages (including obscure ones) to enhance the coverage of search results.
          \item Output format: [Entity1 \textbar Alias1 \textbar Alias2, Entity2 \textbar Alias3 \textbar Alias4 \textbar Alias5].
      \end{itemize}
  \end{adjustwidth}
  \textbf{4.keywords:} Extract a multi-dimensional keyword set from the query, which is used to represent the core intention and retrieval direction of the user.\newline 
  \textbf{5.target\_layer:} According to the characteristics of the query problem, combined with the information of the knowledge hypergraph architecture, predict which layer of the architecture the answer vertices related to the problem are located in (ranging from 1 to 3, type as Int).\newline 
  \textbf{6.matching\_score:} This score (ranging from 1 to 5, type as Int) reflects the degree of relevance between the query answer and the knowledge at the target layer, as well as the complexity of integrating knowledge across different layers during the response.\newline 
     The detailed scoring criteria are as follows:
     \begin{adjustwidth}{0em}{0em}
         \begin{itemize}
             \item 5: Very High, the answer must be present in the target layer, and no other layer knowledge is needed; the query can almost be completely resolved within the target layer without cross-layer reference. Usually, it includes simple and isolated facts, superficial information, or macro themes (queries with macro themes almost only focus on the fourth layer). The knowledge at the target layer can provide about 95\% to 100\% of the content in the answer.
             \item 4: Relatively High, the answer mainly focuses on the target layer, but may still need to call upon a small amount of knowledge from adjacent layers to provide background or supplementary information; most doubts can be resolved within a single domain or layer. The knowledge at the target layer can provide about 80\% to 95\% of the content in the answer.
             \item 3: Medium, the answer mainly relies on the support of the target layer and its upper-layer knowledge, with limited but necessary cross-layer requirements. The knowledge at the target layer can provide about 60\% to 80\% of the content in the answer.
             \item 2: Relatively Low, the answer requires the collaboration of multiple layers to support, especially relying on the adjacent upper and lower layers. Any missing key layer will lead to an incomplete answer. The knowledge at the target layer can provide about 35\% to 60\% of the content in the answer.
             \item 1: Very Low, all or most layers of knowledge must be integrated to provide an accurate answer. The construction of the answer relies on a complete information chain from the raw data to the abstract layer. This is usually related to broader significant events or complex knowledge, and needs to be explored throughout the knowledge base architecture. The knowledge at the target layer can provide about 5\% to 35\% of the content in the answer.
         \end{itemize}
     \end{adjustwidth}
  \textbf{7.semantic\_depth:} Evaluate the complexity, abstraction and reasoning depth of the query to determine whether multi-step reasoning or cross-domain knowledge integration is required and predict the semantic depth of the query (ranging from 1 to 5, type as Int).
      \begin{adjustwidth}{0em}{0em}
         \begin{itemize}
         \item Caution: If you decide to set the semantic depth to 1, do so with caution as it may result in some relevant content being omitted in the subsequent search.
         \end{itemize}
     \end{adjustwidth}
}
\end{myprompt_single}

\subsection{Prompts for Answering}
\label{prompt_answering}
\begin{myprompt_single}{Answering Briefly}
    \textbf{Prompt: }\textit{
---Goal---\newline 
You are a helpful assistant responding to questions about data in the tables provided, and your task is to analyze the data tables provided and the corresponding questions meticulously.\newline 
---Data tables---\newline 
\{info\}\newline 
---Example---\newline 
 -input-\newline 
 Question: When was Neville A. Stanton's employer founded?\newline 
 -output-\newline 
 Thought: The employer of Neville A. Stanton is the University of Southampton. The University of Southampton was founded in 1862.\newline 
 Answer: 1862.\newline 
---Query---\newline 
 -input-\newline 
 Question: \{query\}\newline 
 Your response starts after ``Thought:'', where you will methodically break down the reasoning process, illustrating how you arrive at conclusions.\newline 
 Conclude with ``Answer:'' to present a concise, definitive response, devoid of additional elaborations.\newline 
 -output-
}
\end{myprompt_single}
\begin{myprompt_single}{Answering in Detail}
    \textbf{Prompt: }\textit{
---Goal---\newline 
You are a helpful assistant responding to questions about data in the tables provided, and your task is to analyze the data tables provided and the corresponding questions meticulously.\newline 
---Data tables---\newline 
\{info\}\newline 
---Example---\newline 
 -input-\newline 
 Question: What were the key design features and operational capabilities of the Polar Satellite Launch Vehicle (PSLV) in relation to its development history, and how did these features contribute to its success in launching satellites, particularly in the context of India's evolving space ambitions?\newline 
 -output-\newline 
 The Polar Satellite Launch Vehicle (PSLV) was designed and operated by the Indian Space Research Organisation (ISRO) as a medium-lift launch vehicle specifically developed to place satellites into sun-synchronous orbits...\newline 
---Query---\newline 
 -input-\newline 
 Question: \{query\}\newline 
 The information you are responding with can only be obtained from the provided table.\newline 
 If possible, please provide an accurate and comprehensive answer directly in one or two paragraphs, with a word count of approximately 450 to 600 words.\newline 
 -output-
}
\end{myprompt_single}

\subsection{Prompts for Explanatory Task}
\label{prompt_exp}
\begin{myprompt_single}{Generating One-Hop Question}
    \textbf{Prompt: }\textit{
-Goal-\newline 
You are a professional teacher, and you are now asked to design a question that meets the requirements based on the reference.\newline 
\newline 
-Reference-\newline 
Given the following fragment of a data set:\newline 
\{context\}\newline 
\newline 
-Requirements-\newline 
1. This question should be of the question-and-answer (QA) type, and a precise and comprehensive answer is required.\newline 
2. This question mainly tests the details of the information and knowledge in the reference. Avoid general and macro questions.\newline 
3. The question must not include any conjunctions such as ``specifically'', ``particularly'', ``and'', ``or'', ``and how'', ``and what'' or similar phrases that imply additional inquiries.\newline 
4. The question must focus on a single aspect or detail from the reference, avoiding the combination of multiple inquiries.\newline 
5. Please design a question from the professional perspective and domain factors covered by the reference.\newline 
6. This question needs to be meaningful and difficult, avoiding overly simplistic inquiries.\newline 
7. This question should be based on the complete context, so that the respondent knows what you are asking and doesn't get confused.\newline 
8. State the question directly in a single sentence, without statements like ``How in this reference?'' or ``What about this data set?'' or ``as described in the reference.''\newline 
9. The answer to the question should consist of one or two paragraphs, with a word count ranging from 450 to 500 words; it should contain sufficient contextual information and provide a detailed and thorough explanation of the question.\newline 
\newline 
-Output-\newline 
Question: ...,\newline 
Answer: ...
}
\end{myprompt_single}

\begin{myprompt_single}{Generating Multi-Hop Question}
    \textbf{Prompt: }\textit{
-Goal-\newline 
You are a professional teacher, and your task is to design a single question that contains \{n\} interconnected sub-questions, demonstrating a progressive relationship based on the reference.\newline 
\newline 
-Reference-\newline 
Given the following fragment of a data set:\newline 
\{context\}\newline 
\newline 
-Requirements-\newline 
1. This question should be of the question-and-answer (QA) type, and a precise and comprehensive answer is required.\newline 
2. The question must include \{n\} sub-questions connected by transitional phrases such as ``and'' or ``specifically'', indicating progression.\newline 
3. Focus on testing the details of the information and knowledge in the reference. Avoid general and macro questions.\newline 
4. Design the question from a professional perspective, considering the domain factors covered by the reference.\newline 
5. Ensure the question is meaningful and challenging, avoiding trivial inquiries.\newline 
6. The question should be based on the complete context, ensuring clarity for the respondent.\newline 
7. State the question directly in a single sentence, without introductory phrases like ``How in this reference?'' or ``What about this data set?''.\newline 
8. The answer to the question should consist of one or two paragraphs, with a word count ranging from 500 to 550 (depending on \{n\}) words; it should contain sufficient contextual information and provide a detailed and thorough explanation of the question.\newline 
\newline 
-Output-\newline 
Question: ...,\newline 
Answer: ...
}
\end{myprompt_single}

\begin{myprompt_single}{Evaluating for EM and F1}
    \textbf{Prompt: }\textit{\# Task\newline 
Providing a question and its reference answer, your task is to assess the quality of the given answers.\newline 
-Commonalities identification-\newline 
First, analyze the similarities between the given answer and the reference answer, and summarize these commonalities into several key points.\newline 
-Score-\newline 
The assessment is conducted from three perspectives: \newline 
- **Exact Match** -\newline 
Based on the reference answer, conduct a comprehensive evaluation of the given answer.\newline 
Level   \textbar\ score range \textbar\ description\newline 
Level 1 \textbar\ 0-20 \textbar\ The answer has serious errors and is extremely incomplete, its core facts, data, or conclusions contradict the reference answer, and it omits most essential points; it can hardly answer the question accurately. Inconsistent; basically does not match the reference answer, and is unable to answer the question.\newline 
Level 2 \textbar\ 20-40 \textbar\ The answer has obvious errors and is incomplete; some of its core points do not match the reference answer, and it lacks multiple important pieces of information; the degree of matching with the reference answer and overall accuracy is low, and it can mislead readers.\newline 
Level 3 \textbar\ 40-60 \textbar\ The answer is generally correct but incomplete; its main facts, data, and conclusions are consistent with the reference answer, but there are some minor errors and a few essential points are missing. The facts are correct, but there are still some data, details, or conclusions that do not match the correct answer.\newline 
Level 4 \textbar\ 60-80 \textbar\ The answer is basically correct and relatively complete; its majority of facts, data, and conclusions are consistent with the reference answer, and only a few details or expressions have minor deviations, with very few missing points; it is matched closely with the reference answer and can accurately answer the question.\newline 
Level 5 \textbar\ 80-100 \textbar\ The answer is completely correct and complete: all core facts, data, and conclusions are consistent with the reference answer, there are no factual errors, and it covers all the required points of the question; it is matched completely with the reference answer so that can be regarded as the benchmark for an accurate answer to the question.\newline 
- **Recall** -\newline 
Measure the degree of information coverage of the given answer relative to the reference answer; the more comprehensive the coverage, the higher the score; the more omissions, the lower the score.\newline 
Level   \textbar\ score range \textbar\ description\newline 
Level 1 \textbar\ 0-20 \textbar\ The given answer fails to cover any key facts, entities or conclusions of the reference answer; a large amount of core information is missing.\newline 
Level 2 \textbar\ 20-40 \textbar\ The given answer only touches upon a few marginal details of the reference answer; the main facts and key arguments are still largely omitted.\newline 
Level 3 \textbar\ 40-60 \textbar\ The given answer covers the main part of the reference answer, but there are still significant core facts, details or sub-arguments missing or omitted.\newline 
Level 4 \textbar\ 60-80 \textbar\ The given answer covers the majority of the key information and core arguments of the reference answer, with only a few minor details or sub-arguments missing.\newline 
Level 5 \textbar\ 80-100 \textbar\ The given answer fully covers all the key facts, entities, logical chains and details of the reference answer, without any substantive omissions.\newline 
- **Precision** -\newline 
Measures the degree of consistency between the given answer and the reference answer in terms of semantics, facts, and details; the higher the similarity and fewer the interfering factors, the higher the score; the greater the contrast and the more redundant or incorrect elements, the lower the score.\newline 
Level 1 \textbar\ 0-20 \textbar\ The given answer has almost no intersection with the reference answer, contains a large number of factual errors, opposite statements, or irrelevant noise, and can be regarded as irrelevant to the question.\newline 
Level 2 \textbar\ 20-40 \textbar\ The given answer shares only a few common points with the reference answer, but the core facts, key entities, or conclusions significantly differ from those of the reference answer.\newline 
Level 3 \textbar\ 40-60 \textbar\ The given main framework or some key information of the answer is consistent with the reference answer, but there are still obvious factual deviations, omissions, or redundant details.\newline 
Level 4 \textbar\ 60-80 \textbar\ The majority of the key facts, entities, and logic of the given answer closely match those of the reference answer, with only a few details or expressions having slight differences or redundancies.\newline 
Level 5 \textbar\ 80-100 \textbar\ The given answer is almost one-to-one corresponding to the reference answer in terms of facts, entities, logic, and details; synonyms can be used, and there is no additional noise or contradictions.
\newline
\# Question\newline
\{query\}\newline
\# Reference Answer\newline
\{gold\_answer\}\newline
\# Answer that Requires Judgment\newline
\{pre\_answer\}\newline
Evaluate the answer using the criteria above: You need to provide the level and explanation for the answer according to the indicator descriptions, and then give the score within the rating range of that level.\newline
Output your evaluation in the following JSON format:\newline
\{\{\newline
    ``Commonalities'':``The common contents between the given answer and the reference answer obtained through analysis''\newline
    ``Exact Match'': \{\{\newline
        ``Explanation'': ``Provide explanation here''\newline
        ``Level'': ``A level range 1 to 5''  \# This should be a single number, not a range\newline
        ``Score'': ``A value range 0 to 100''  \# This should be a single float number that is precise to two decimal places, not a range\newline
    \}\},\newline
    ``Recall'': \{\{\newline
        ``Explanation'': ``Provide explanation here''\newline
        ``Level'': ``A level range 1 to 5''  \# This should be a single number, not a range\newline
        ``Score'': ``A value range 0 to 100''  \# This should be a single float number that is precise to two decimal places, not a range\newline
    \}\},\newline
    ``Precision'': \{\{\newline
        ``Explanation'': ``Provide explanation here''\newline
        ``Level'': ``A level range 1 to 5''  \# This should be a single number, not a range\newline
        ``Score'': ``A value range 0 to 100''  \# This should be a single float number that is precise to two decimal places, not a range\newline
    \}\},\newline
\}\}
}
\end{myprompt_single}

\end{document}